\documentclass{aastex701}

\usepackage{amsmath}
\usepackage{bm}
\usepackage{booktabs}
\usepackage{multirow}

\shorttitle{Proton Acceleration in SMBH Coronae}
\shortauthors{Ly et al.}
\journalinfo{The Astrophysical Journal, 1004:27 (14pp), 2026 June 10}
\received{2026 January 5}
\revised{2026 April 16}
\accepted{2026 April 18}
\published{2026 June 3}

\begin{document}

\title{Proton Acceleration by Collisionless Shocks in Supermassive Black Hole Coronae: Implications for High-energy Neutrinos}

\author[0009-0009-6218-5446]{Minh Nhat Ly}
\affiliation{Department of Physics, Graduate School of Science, The University of Osaka, 1-1 Machikaneyama, Toyonaka, Osaka 560-0043, Japan}
\affiliation{Institute of Laser Engineering, The University of Osaka, 2-6 Yamadaoka, Suita, Osaka 565-0871, Japan}
\email[show]{minhly.phys@gmail.com}

\author[0000-0002-7272-1136]{Yoshiyuki Inoue}
\affiliation{College of Systems Engineering and Science, Shibaura Institute of Technology, 307 Fukasaku, Minuma-ku, Saitama City, Saitama 337-8570, Japan}
\affiliation{Department of Earth and Space Science, Graduate School of Science, The University of Osaka, 1-1 Machikaneyama, Toyonaka, Osaka 560-0043, Japan}
\affiliation{Interdisciplinary Theoretical \& Mathematical Science Center (iTHEMS), RIKEN, 2-1 Hirosawa, Wako, Saitama, 351-0198, Japan}
\affiliation{Kavli Institute for the Physics and Mathematics of the Universe (WPI), UTIAS, The University of Tokyo, 5-1-5 Kashiwanoha, Kashiwa, Chiba 277-8583, Japan}
\email[]{yinoue@astro-osaka.jp}

\author[0000-0003-2093-7100]{Yasuhiko Sentoku}
\affiliation{Institute of Laser Engineering, The University of Osaka, 2-6 Yamadaoka, Suita, Osaka 565-0871, Japan}
\email[]{sentoku.yasuhiko.ile@osaka-u.ac.jp}

\author[0000-0001-9106-3856]{Takayoshi Sano}
\affiliation{Institute of Laser Engineering, The University of Osaka, 2-6 Yamadaoka, Suita, Osaka 565-0871, Japan}
\email[]{sano.takayoshi.ile@osaka-u.ac.jp}

\begin{abstract}

Recent observations by the IceCube Neutrino Observatory have revealed a significant excess of high-energy neutrinos from nearby Seyfert galaxies, such as NGC~1068, without a corresponding flux of high-energy gamma-rays. 
This suggests that neutrinos are produced via hadronic interactions in a region opaque to gamma-rays, likely a hot corona surrounding the central supermassive black hole. However, the mechanism responsible for accelerating the parent protons to the required energies ($\sim 100$ TeV) remains an open question. 
In this study, we investigate diffusive shock acceleration (DSA) in active galactic nucleus (AGN) coronae using a suite of one-dimensional particle-in-cell simulations spanning a broad range of plasma parameters. 
We find that DSA is a robust and efficient mechanism for proton acceleration, consistently channeling approximately 10\% of the shock's kinetic energy into nonthermal ions, even for shocks with sonic Mach numbers as low as $ M_s \approx 2$. 
In contrast, the efficiency of electron acceleration is highly variable and less efficient ($<1\%$) in our parameter survey.
These findings provide strong, first-principles support for the hadronic models of neutrino production in AGN, and offer quantitative constraints that can explain the observed gamma-ray deficit.

\end{abstract}


\section{Introduction} \label{sec:intro}

High-energy astrophysical neutrinos are produced in hadronuclear (\(pp\)) or photopion (\(p\gamma\)) interactions of high-energy cosmic rays (CRs).
These neutrinos are usually accompanied by a comparable gamma-ray flux according to standard hadronic scenarios, \(pp\) and \(p\gamma\)~\citep[e.g.,][]{kelner2006EnergySpectraGamma, kelner2008EnergySpectraGamma}.
Recently, the IceCube Collaboration reported TeV neutrinos from nearby, radio-quiet (RQ) active galactic nuclei (AGNs), most clearly the Seyfert galaxy NGC~1068~\citep{icecubecollaboration*+2022EvidenceNeutrinoEmission, abbasi_evidence_2025}.
However, the observed gamma-ray fluxes are lower than expected, compared to the neutrino fluxes, which cannot be explained by a simple hadronic picture.
This discrepancy points to a gamma-ray-opaque production site for the neutrinos, although other possibilities are still being discussed~\citep{yasuda2024NeutrinosGammaRays}.

A natural candidate is a hot corona surrounding the central supermassive black hole \citep[SMBH; ][]{inoue2019HighenergyParticlesAccretion, murase2020HiddenCoresActive}.
A black hole (BH) corona is a compact, hot plasma with electron temperature \(T_{\rm e}\approx 10^{9}\,\mathrm{K}\) and a characteristic size of about \(100\,R_{\rm g}\), where \(R_{\rm g}\) is the gravitational radius.
Protons would be decoupled from electrons and have temperatures $T_{\rm i} \gg T_{\rm e}$, due to the collisionless nature of the environment \citep[e.g.,][]{kawabata2010PASJ...62..621K}.
In such a region, Comptonized X-ray photons can efficiently attenuate GeV to TeV gamma-rays through \(\gamma\gamma\) pair production, while neutrinos can escape.
In order to explain the IceCube signal, the corona must host parent CRs accelerated to at least \(\sim 100\)~TeV so that secondary pions can produce the observed neutrinos while the associated gamma-rays are suppressed~\citep{icecubecollaboration*+2022EvidenceNeutrinoEmission}.

However, the dominant particle acceleration mechanism in SMBH coronae is not yet established.
Proposed scenarios include diffusive shock acceleration  \citep[DSA;][]{inoue2019HighenergyParticlesAccretion, yoshiyuki_inoue_origin_2020, Inoue2022arXiv220702097I}, stochastic acceleration \citep[SA;][]{murase2020HiddenCoresActive, kheirandish2021HighenergyNeutrinosMagnetized, fiorillo2024MagnetizedStronglyTurbulenta, mbarek2024InterplayAcceleratedProtons, Groselj2026arXiv260100518G}, magnetic reconnection \citep{fiorillo2023TeVNeutrinosHard, karavola2024NeutrinoPairCreation}, and hybrid cases where these processes act together (see, e.g., \citealt{lemoine_neutrinos_2025}, who invoked SA combined with reacceleration in the jet).

Compared to other mechanisms, recent observational results favor DSA.
In particular, millimeter observations of nearby Seyferts point to coronal magnetic field strengths of roughly \(10\)--\(100\,\mathrm{G}\)~(\citealt{inoue2018DetectionCoronalMagnetica, yoshiyuki_inoue_origin_2020, Michiyama2023PASJ...75..874M, Michiyama2024ApJ...965...68M, Shablovinskaya2024A&A...690A.232S, palacio_millimeter_2025, Jana2025A&A...699A..62J, mutie_consistent_2025}, but see also \citealt{Hankla2025arXiv251201662H}) corresponding to Alfv\'en speeds of $v_A/c \sim 10^{-2}-10^{-3}$.
Such weak field conditions are generally unfavorable for efficient SA or magnetic reconnection. 
Nonetheless, the viability of each mechanism under realistic coronal conditions remains a subject of active debate, partly due to the limited first-principles constraints tailored to SMBH coronae.

The largest uncertainty for DSA lies in the efficiency of its particle acceleration within coronal plasmas.
The hot coronal environment likely results in low sonic Mach number shocks ($M_s$ of order a few), which are traditionally considered to produce weak shocks and thus would be unable to accelerate particles efficiently~\citep{ha_proton_2018, kang_electron_2019, vanmarle_influence_2020}.
Moreover, the parameter regime for coronal shocks has yet to be explored consistently for both electron and ion dynamics.
In particular, the behavior of nonequilibrium, two-temperature plasmas ($T_i \neq T_e$) and the trans-relativistic shock speeds remain poorly understood.

To resolve uncertainties and provide better constraints on DSA as a leading candidate for proton acceleration, we investigate collisionless shocks by employing first-principles, particle-in-cell (PIC) simulations tuned to BH coronal parameters.
We survey quasi-parallel shocks across the ranges of sonic Mach number, Alfvénic Mach number, shock velocity, and ion-to-electron temperature ratio, motivated by observational constraints for Seyferts~\citep{inoue2019HighenergyParticlesAccretion}.

This paper is organized as follows. In Section~\ref{sec:condition_BHCS}, we derive the shock conditions expected in BH coronae.
The PIC simulation setup is then described in detail in Section~\ref{sec:setup_PIC}. 
In Section~\ref{sec:results}, we present our main findings, beginning with the shock structure and a unified picture of particle acceleration from a representative simulation (Section~\ref{subsec:typical_sim}). 
We then examine the proton acceleration rate (Section~\ref{subsec:proton_acceleration}) and the proton and electron acceleration efficiencies across our parameter survey (Section~\ref{subsec:energy_partition}).
In the discussion (Section~\ref{sec:discussion}), we explore the implications of our results for neutrino production in BH coronae, focusing on the well-known NGC~1068. 
We also briefly address electron acceleration in Section~\ref{subsection:disscus_electron}. Finally, we summarize our conclusions in Section~\ref{sec:conclusion}.

\section{Shock conditions in BH coronae} \label{sec:condition_BHCS}

\begin{figure}[t]
     \centering
     \includegraphics[width=0.95\textwidth]{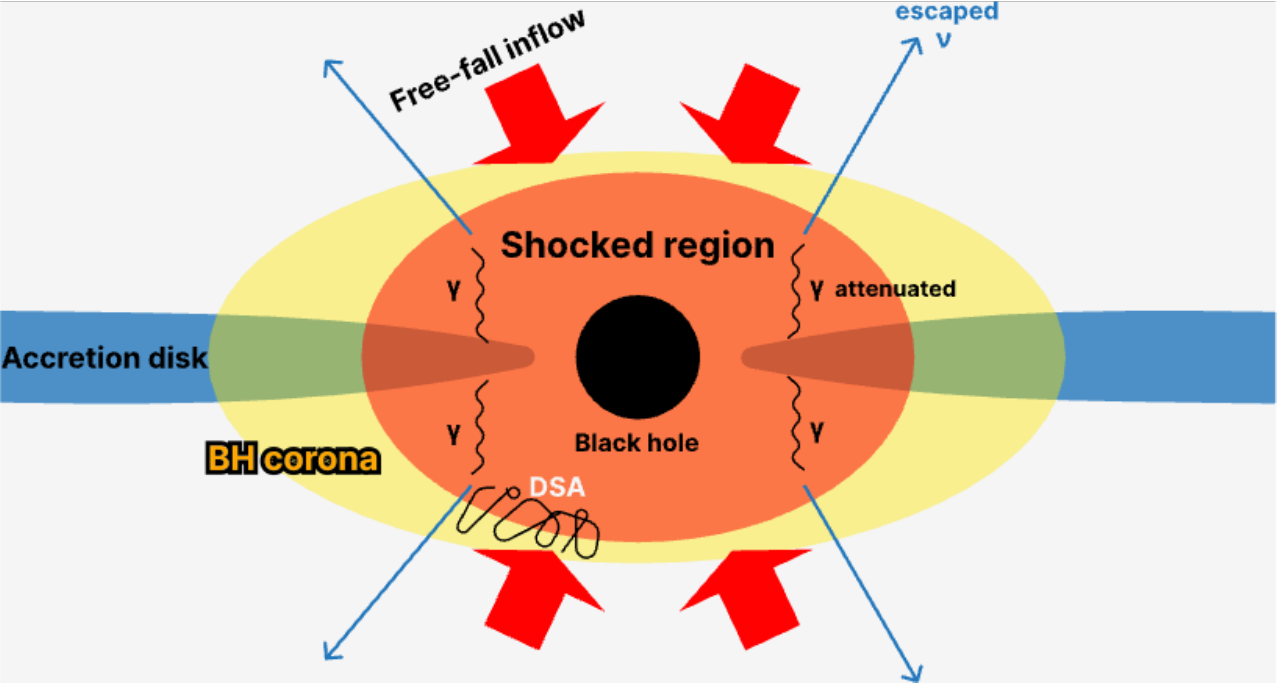}
     \caption{\label{fig:schematic}
     Schematic illustration of an accretion shock formed by the infalling flow in the hot BH corona. Protons are accelerated to high energies and produce high-energy neutrinos through hadronic \(pp\) and \(p\gamma\) interactions. Neutrinos can escape from the system and be observed by IceCube, while the accompanying gamma-rays will be attenuated by X-ray photons.
     }
\end{figure}

Here, the corona of an SMBH of mass $M_{\rm BH}$ is considered as a hot, collisionless plasma region extending to $R_c \lesssim 100\,R_g$, where $R_g = G M_{\rm BH}/c^2$ is the gravitational radius.
We consider a scenario in which shocks form within the BH corona (see Figure~\ref{fig:schematic}). 
Such shocks can naturally arise from infalling material associated with the accretion flow or from failed winds~\citep{inoue2019HighenergyParticlesAccretion, yoshiyuki_inoue_origin_2020, Inoue2022arXiv220702097I}.
Shocks may also be triggered by material falling in from larger radii, for example, from broad-line region clouds~\citep{muller2020RadiationImpactBroadline, muller2022NonthermalEmissionFallback, sotomayor2022NonthermalRadiationCentral}. 
In this case, the fallback flow is likely much colder than the BH corona, which we will refer to as the cold inflow scenario.

Kinetic studies of shock-accelerated particles in collisionless plasmas have identified shock sonic and Alfv\'enic Mach number, $M_s$ and $M_A$, as the most important quantities to determine characteristics of particle acceleration.
In this section, we will derive these shock parameters from the conditions in BH coronae.
Our goal is to establish a fiducial range for the shock parameters consistent with multi-messenger observations, thereby providing a physical basis for modeling shock-driven proton acceleration and subsequent neutrino production.
For this work, we assume the plasma is primarily composed of ions and electrons, where the fraction of positrons is negligible.
We treat ions as protons, which constitute the majority of baryonic matter.

For our fiducial model, we assume the shock velocity, $v_{\rm sh}$, is on the order of the free-fall velocity of infalling material, $v_{\rm ff} = \sqrt{2 G M_{\rm BH}/R_c}$.
This yields the following scaling for shock speed:
\begin{equation}
    \frac{v_{\rm sh}}{c}
    \simeq 0.14 \left( \frac{r_c}{100} \right)^{-1/2}\;,
    \label{eq:v_sh}
\end{equation}
where $r_c = R_c/R_g$ is the coronal radius in the unit of the gravitational radius.

The sonic Mach number $M_s = v_{\rm sh}/c_s$ depends on the plasma sound speed, $c_s=\sqrt{\gamma_{\rm ad}(T_i + T_e)/m_i}$, where $T_i$ and $T_e$ denote the ion (proton) and electron temperatures, respectively, and $m_i = m_p$ is the rest mass of the proton.
We also adopt an adiabatic index of $\gamma_{\rm ad}=5/3$.
While the electron temperature can be constrained by the cutoff energy of the X-ray spectrum to $T_e \sim 100~\text{keV}$~\citep[e.g.,][]{fabian_properties_2015, laha_x-ray_2025}, the ion temperature is poorly constrained by observations.
In BH hot coronae, inefficient Coulomb coupling allows ions to become significantly hotter than electrons ($T_i \gg T_e$).
A common assumption of $T_i \sim 10\,T_e \sim 1~\text{MeV}$ leads to
\begin{equation}
    M_s = \frac{v_{\rm sh}}{c_s} \simeq 3.4 \left( \frac{r_c}{100} \right)^{-1/2}
          \left( \frac{T_i}{1\,\text{MeV}} \right)^{-1/2}\;.
    \label{eq:Ms}
\end{equation}
Alternatively, a more physically motivated estimate for $T_i$ comes from advection-dominated accretion flow (ADAF) models, which predict $T_i \simeq G M_{\rm BH} m_i/(6 R_c) \sim 1.6 (r_c/100)^{-1}~\text{MeV}$~\citep{yuan2014HotAccretionFlows, inoue2024UpperLimitCoronal}.
This temperature profile yields a slight lower Mach number of $M_s \simeq 2.7$ independent of other parameters.
Based on the estimation, shocks in the BH corona are likely to have low sonic Mach numbers, although higher sonic Mach numbers (up to a few tens) could be realized in the cold inflow picture.

The Alfv\'enic Mach number, $M_A = v_{\rm sh}/v_A$, depends on the Alfv\'en speed, $v_A=B_0/\sqrt{4\pi n_0 m_i}$, which requires estimates for the upstream plasma density $n_0$ and magnetic field $B_0$.
We estimate the electron density $n_e=n_0$ from Thomson scattering opacity, $\tau_T = n_e \sigma_T R_c \sim 1$~\citep{inoue2019HighenergyParticlesAccretion, padovani2024HighenergyNeutrinosVicinity}.
The coronal magnetic field $B_0$ remains a significant source of uncertainty.
Millimeter emissions from electron synchrotron radiation are suggested as a method to estimate $B_0$, which constrains $B_0 \sim O(10-100~\text{G})$~\citep{inoue2018DetectionCoronalMagnetica, palacio_millimeter_2025}.
For example, the value estimated for NGC~1068 is $ B_0\simeq 150~\text{G}$~\citep{palacio_millimeter_2025, mutie_consistent_2025}. 
Hence, we adopt $B_0 = 100~\text{G}$ as the fiducial value for this study.
The Alfv\'en speed and the Alfv\'enic Mach number are then given by
\begin{equation}
    \frac{v_A}{c} \simeq 7.2\times10^{-3} \left( \frac{r_c}{100} \right)^{1/2}
                                     \left( \frac{M_{\rm BH}/M_{\odot}}{10^7} \right)^{1/2}
                                     \left( \frac{B_0}{100\,\text{G}} \right)\;,
    \label{eq:Va}
\end{equation}
\begin{equation}
    M_A = \frac{v_{\rm sh}}{v_A} \simeq 20 \left( \frac{r_c}{100} \right)^{-1}
                                           \left( \frac{M_{\rm BH}/M_{\odot}}{10^7} \right)^{-1/2}
                                           \left( \frac{B_0}{100\,\text{G}} \right)^{-1}\;.
    \label{eq:Ma}
\end{equation}
It is worth noting that the estimate of magnetic field strength is still under debate. 
Other acceleration models, such as SA or reconnection models, require stronger fields \citep[$B \gg 10^3$ G, see e.g., ][]{murase2022HiddenHeartsNeutrino, das2024RevealingProductionMechanism, fiorillo2023TeVNeutrinosHard}.

In summary, our theoretical estimates suggest that shocks in the coronae of RQ AGNs are likely to be trans-relativistic ($v_{\rm sh}/c \sim 0.3$), with low sonic Mach numbers ($M_s \sim 2-10$) and moderate Alfv\'en Mach numbers ($M_A \sim 30$).
These values are subject to considerable uncertainty; hence, we aim to systematically survey shocks with a wide range of parameters (see details in Table~\ref{tab:sim_params}).
We will focus on the effects that are less well-known from previous collisionless shock studies, like the trans-relativistic shock regime ($v_{\rm sh}/c \sim 0.3$) and initial ion-to-electron temperature imbalance ($T_i/T_e$).

\section{PIC simulation setup} \label{sec:setup_PIC}

\begin{table}[h!]
\centering
\caption{Simulation Parameter Survey} 
\label{tab:sim_params}
\begin{tabular}{l c c c c c c l}
\toprule
\textbf{Run ID} & \textbf{$v_{\rm sh}/c$} & \textbf{$M_s$} & \textbf{$M_A$} & \textbf{$T_i/T_e$} & \textbf{$T_e/(m_i c^2)$} & \textbf{$T_i/(m_i c^2)$} & \textbf{Purpose} \\
\midrule
T10 (Ref)   & 0.33  & 8  & 26 & 10    & $9.281\times 10^{-5}$ & $9.281\times 10^{-4}$ & Reference case \\
\midrule
T1          & 0.33  & 8  & 26 & 1     & $5.105\times 10^{-4}$ & $5.105\times 10^{-4}$ & \multirow{3}{*}{Different temperature ratios ($T_i/T_e$)} \\
T50         & 0.33  & 8  & 26 & 50    & $2.002\times 10^{-5}$ & $1.001\times 10^{-3}$ & \\
T0.01       & 0.33  & 8  & 26 & 0.01  & $1.011\times 10^{-3}$ & $1.011\times 10^{-5}$ & \\
\midrule
S4T1        & 0.33  & 4  & 26 & 1     & $2.042\times 10^{-3}$ & $2.042\times 10^{-3}$ & \multirow{3}{*}{Different temperature ratios with $M_s = 4$} \\
S4T10       & 0.33  & 4  & 26 & 10    & $3.713\times 10^{-4}$ & $3.713\times 10^{-3}$ & \\
S4T50       & 0.33  & 4  & 26 & 50    & $8.007\times 10^{-5}$ & $4.004\times 10^{-3}$ & \\
\midrule
S2T1        & 0.33  & 2  & 26 & 1     & $8.168\times 10^{-3}$ & $8.168\times 10^{-3}$ & \multirow{2}{*}{Different temperature ratios with $M_s = 2$} \\
S2T10       & 0.33  & 2  & 26 & 10    & $1.485\times 10^{-3}$ & $1.485\times 10^{-2}$ & \\
\midrule
S20T1       & 0.33  & 20 & 26 & 1     & $8.168\times 10^{-5}$ & $8.168\times 10^{-5}$ & Higher $M_s$ with $T_i=T_e$ \\
S30T0.01    & 0.33  & 30 & 26 & 0.01  & $7.188\times 10^{-5}$ & $7.188\times 10^{-7}$ & Higher $M_s$ with hotter electrons \\
\midrule
A4          & 0.33  & 8  & 4  & 10    & $9.281\times 10^{-5}$ & $9.281\times 10^{-4}$ & \multirow{3}{*}{Different Alfv\'en Mach numbers ($M_A$)} \\
A8          & 0.33  & 8  & 8  & 10    & $9.281\times 10^{-5}$ & $9.281\times 10^{-4}$ & \\
A46         & 0.33  & 8  & 46 & 10    & $9.281\times 10^{-5}$ & $9.281\times 10^{-4}$ & \\
\midrule
V0.125      & 0.167 & 8  & 26 & 10    & $2.377\times 10^{-5}$ & $2.377\times 10^{-4}$ & \multirow{2}{*}{Different shock velocities ($v_{\rm sh}$)} \\
V0.5        & 0.66  & 8  & 26 & 10    & $3.713\times 10^{-4}$ & $3.713\times 10^{-3}$ & \\
\bottomrule
\end{tabular}
\end{table}

To reveal a first-principles-based, self-consistent description of particle acceleration under the physical conditions of BH coronae, we employ the PIC method.
A series of large-scale 1D3V (one spatial dimension and three velocity dimensions) PIC simulations using the SMILEI code \citep{derouillat_smilei_2018} are adopted to survey the parameter space of the shocks in BH coronae.

Suppose a uniform plasma with the density $n_i = n_e$ occupies the region $x>0$.
The plasma is initialized with a velocity $-v_{\rm pt}$ drifting toward a reflective boundary at $x=0$, where the incoming and reflected plasmas interact and launch a shock toward the $+x$ direction.
Shocks modeled this way coincide with the shocks' downstream rest frame, similar to previous studies~\citep{park2015SimultaneousAccelerationProtons, shalaby2024EnergyDissipationStrong}.
We introduce the shock compression ratio $\mathcal{R}$. 
In the upstream rest frame, the shock velocity is given by
$v_{\rm sh} = v_{\rm pt} \mathcal{R}/(\mathcal{R}-1)$.
For a strong shock with $\mathcal{R}\approx 4$, this expression reduces to
$v_{\rm sh} \simeq 4v_{\rm pt}/3$.
The magnetic field is initialized in the \(x-y\) plane $\bm{B} = B_x\hat{\bm x}+B_y\hat{\bm y}$, where $B_x=B_0 \cos{\theta}$, $B_y=B_0 \sin{\theta}$, and $B_0$ denotes the field strength.
In this study, we focus on quasi-parallel shocks with a fixed magnetic obliquity $\theta = 30^\circ$.

Simulation parameters are summarized in Table~\ref{tab:sim_params}.
Electrons and protons are initialized with relativistic Maxwell-J\"uttner distributions of $T_{e}$ and $T_{i}$, respectively.
We use a reduced ion-to-electron mass ratio of $m_R = m_i/m_e=100$ to make the simulations computationally feasible but large enough for standard scale-separation between ions and electrons.
Accordingly, the physical parameters ($T_i, T_e, B_0$) are scaled to preserve the key dimensionless parameters ($M_s, M_A, T_i/T_e$) and retain the characteristic of shock plasma physics.
Note that, even if the Mach number is the same, $T_i$ and $T_e$ can vary significantly depending on the temperature ratio $T_i/T_e$ (see Table~\ref{tab:sim_params}).

We resolved most of the simulations with spatial resolution $\Delta x = 0.1~d_e$ and temporal resolution $\Delta t = 0.04~\omega_{\rm pe}^{-1}$, where $\omega_{\rm pe} = \sqrt{4\pi n_0e^2/m_e}$ and $d_e = c/\omega_{\rm pe}$ are in turn the upstream plasma frequency and electron's inertial length.
The only exception is run A4, where we adopted $\Delta x = 0.025~d_e$ and $\Delta t = 0.02~\omega_{\rm pe}^{-1}$ to resolve the electron Larmor radius $\rho_e = m_e v_{\rm pt} c/(eB_0) \simeq 0.3~d_e$ with proper resolution and ensure the conservation of energy.
In the simulations, we use 150 particles per cell per species.
We find that it is more convenient for the analysis of shock structure and ion acceleration to express time in units of the ion gyrofrequency, $\Omega_{\rm ci} = eB_0/(m_i c)$ (Sections~\ref{subsec:typical_sim} and~\ref{subsec:proton_acceleration}). We therefore convert all times to this normalization, using
\begin{equation}
\Omega_{\rm ci} = \omega_{\rm pe}\,\frac{v_{\rm sh}/c}{M_A\sqrt{m_R}} \, .
\end{equation}
All simulations are evolved for at least $t_{\rm end} \gtrsim 400\,\Omega_{\rm ci}^{-1}$, ensuring that the relevant electromagnetic instabilities reach saturation and that a well-developed nonthermal power-law tail emerges in the particle distribution.
We use a sufficiently large simulation box size $L_x$ so that reflected particles and those undergoing acceleration remain within the receding plasma for the entire duration up to $t_{\rm end}$.
The simulation box has length $L_x = 180224~d_e$ for our typical case, with the largest case (run V0.5) reaching $L_x = 294912~d_e$.

\section{Results} \label{sec:results}
\subsection{Quasi-parallel shock structures} \label{subsec:typical_sim}

First, let us examine the results of our reference simulation (run T10), in which the plasma parameters are chosen to be $M_s = 8$, $M_A = 26$, $v_{\rm pt}/c = 0.25$, and $T_i/T_e = 10$.

\subsubsection{Shock structure} \label{subsubsec:shock_structure}

\begin{figure}[t]
     \centering
     \includegraphics[width=0.45\textwidth]{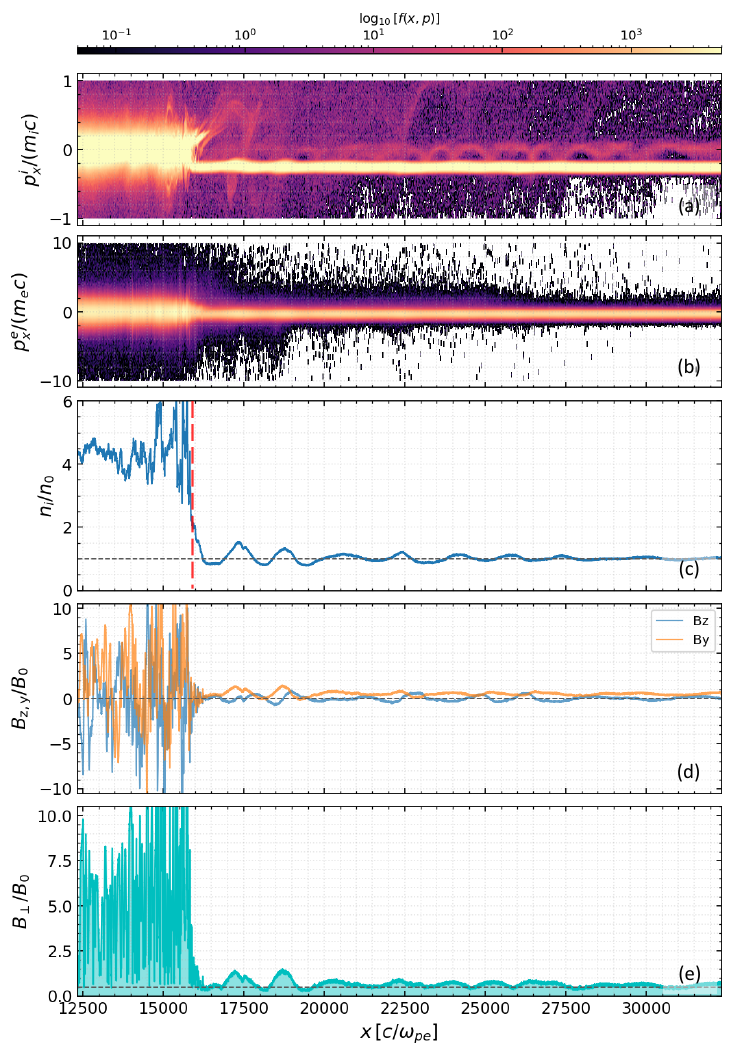}
     \caption{\label{fig:shock_structure}
     Snapshot of shock structures in the fiducial model (run T10) at $t \approx 256 ~\Omega_{\rm ci}^{-1}$. Panels (a) and (b) show the position--momentum (\(x-p_x\)) phase-space distributions for ions and electrons. Panel (c) shows the ion density profile, with the shock position ($x_{\rm sh}$) marked by the red dashed line. Panels (d) and (e) show the transverse magnetic field components ($B_y, B_z$) and the total transverse amplitude ($B_{\perp}$), indicating significant amplification over the initial value [black dashed line in (e)].
     }
\end{figure}

Figure~\ref{fig:shock_structure} displays a snapshot from the reference simulation (run T10) in a quasi-steady evolution stage at $t \approx 256~\Omega_{\rm ci}^{-1}$ ($t=2\times10^5~\omega_{\rm pe}^{-1}$).
The shock front is located at $x_{\rm sh} \approx 16000~d_e$ [vertical red dashed line in Figure~\ref{fig:shock_structure}(c)], marking the boundary between the upstream (unshocked) and downstream (shocked) regions.
The ion density shows a strong compression across the shock, reaching more than four times the upstream density on average ($\mathcal{R} \gtrsim 4$), which signals the formation of a strong shock based on the Rankine-Hugoniot jump conditions.
We define $x_{\rm sh}$ throughout this study as the position of the sharp ion density ramp, where $x \leq x_{\rm sh}$ corresponds to the shock's downstream and $x \geq x_{\rm sh}$ to the upstream region.

In Figures~\ref{fig:shock_structure}(a) and (b), we show the $x-p_x$ phase space distribution of ions and electrons, respectively.
The ion phase space [Figure~\ref{fig:shock_structure}(a)] provides direct evidence for ion acceleration, showing an energetic backstreaming population of ions comprising a shock-reflected and accelerated population, both moving away from the shock into the upstream region.
This streaming ion population excites magnetic fluctuation modes in the upstream [see Figure~\ref{fig:shock_structure}(e)], providing the necessary scattering environment for electrons and ions to complete multiple shock-crossing cycles characteristic of the DSA process.
In addition, the backstreaming ions also generate a net current, which induces a compensating return current in the background electrons.
As shown in the electron phase space [Figure~\ref{fig:shock_structure}(b)], this interaction leads to significant electron preheating in the upstream region near the shock front ($x_{\mathrm{\rm sh}} < x \lesssim 2.5\times10^{4}\,d_e$).
This preheating process has recently been shown to play a crucial role in subsequent electron acceleration in quasi-parallel collisionless shocks~\citep{gupta2024ReturnCurrentsCollisionless, gupta2025SpeeddependentThresholdElectron}.

The transverse magnetic field components, $B_y$ and $B_z$, are shown in Figure~\ref{fig:shock_structure}(d), while their combined magnitude, $B_\perp = \sqrt{B_y^2 + B_z^2}$, is presented in Figure~\ref{fig:shock_structure}(e).
The transverse magnetic amplitude, $B_\perp$, is amplified by approximately a factor $\sim 2$ relative to its 
initial upstream value.
As discussed above, this amplification and the associated fluctuations originate from the interaction between the backstreaming ions and the upstream plasma, the detailed mechanism of which will be investigated in a later section.

\subsubsection{Particle spectrum and acceleration trajectory} \label{subsubsec:spectrum_trajectory}

\begin{figure}[t]
     \centering
     \includegraphics[width=0.45\textwidth]{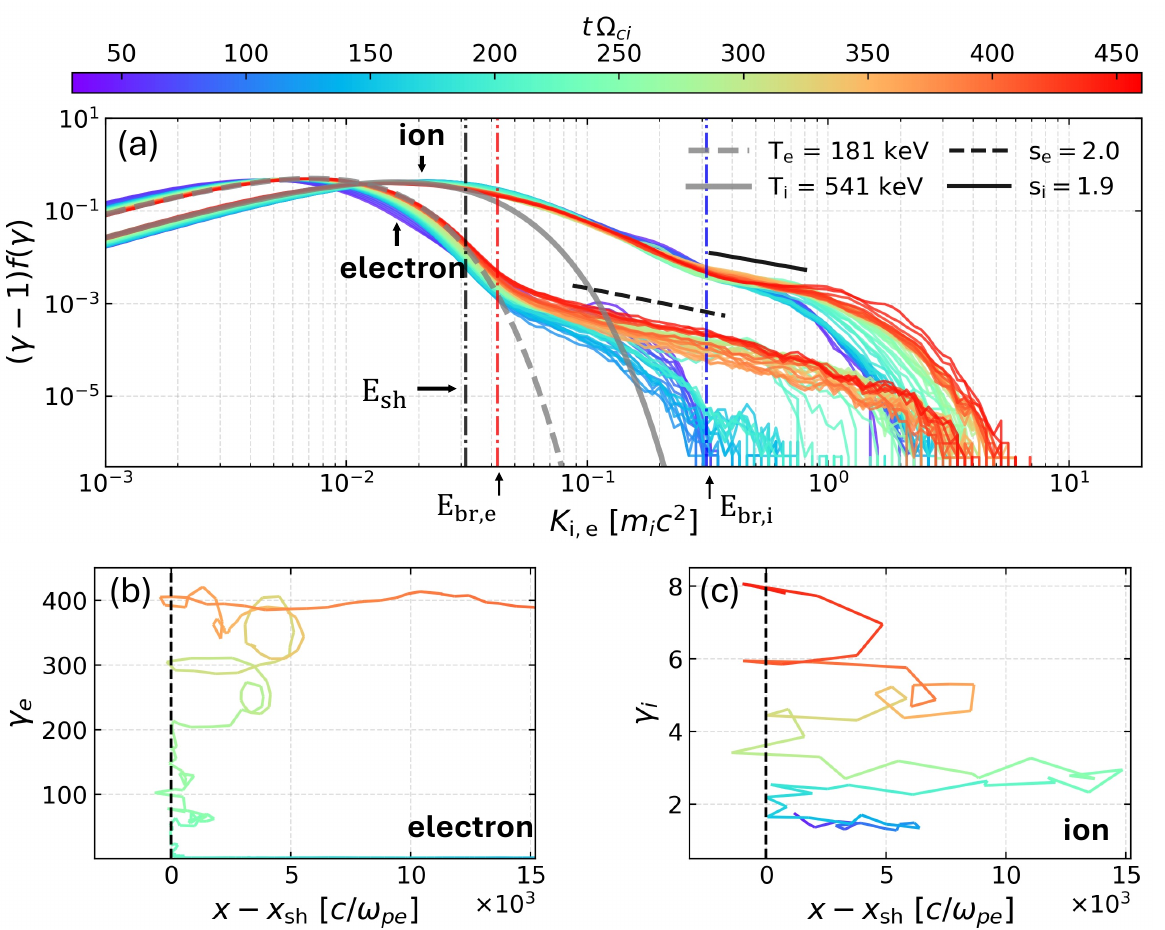}
     \caption{\label{fig:spec_track}
     Signatures of particle acceleration, with the color scale indicating time in units of $t\Omega_{\rm ci}$. Panel (a) shows the evolution of the ion and electron energy spectra in the shock downstream. Both species exhibit two distinct components: a thermal bulk and a developing nonthermal power-law tail. Panels (b) and (c) display the trajectories of a typical accelerating electron and ion, illustrating the DSA process.
     }
\end{figure}

To illustrate how the energy is partitioned between species, each spectrum is plotted as $(\gamma - 1)f(\gamma)$ against the kinetic energy $K = (\gamma - 1)\mu$, where $\mu = m_{i,e}/m_i$ denotes the normalized particle mass.  
The upper panel (a) of Figure~\ref{fig:spec_track} shows the temporal evolution of the ion and electron energy spectra in the downstream region of the shock, spanning $x_{\rm sh} - 6500~d_e<x< x_{\rm sh} - 1500~d_e$.  
Unless stated otherwise, this spatial range is adopted for all analyses of the downstream particle spectra.  
The color gradient represents the simulation time in units of $t\Omega_{\rm ci}$ shared across all panels in Figure~\ref{fig:spec_track}.
  
Both species exhibit clear power-law tails extending from their thermal populations, which can be fitted using a relativistic Maxwellian distribution,
\begin{equation}
    f_{\rm MB}(\gamma) = 
    \frac{\gamma \sqrt{\gamma^2 - 1}}{\Theta K_2(1/\Theta)} 
    \exp\!\left(-\frac{\gamma}{\Theta}\right),
    \label{eq:MB}
\end{equation}
where $\Theta = T/(mc^2)$ is the dimensionless temperature and $K_2$ is the modified Bessel function of the second kind.  
The peak of the ion spectrum coincides with the kinetic energy of the incoming flow, $E_{\rm sh} = \frac{1}{2} m_i v_{\rm pt}^2$ [vertical black dash--dotted line in Figure~\ref{fig:spec_track}(a)], indicating that the bulk of the injected energy is deposited into the thermal ions.  

The break Lorentz factors $\gamma_{\rm br}$, marking the separation point between thermal and nonthermal populations, are shown by red and blue dash--dotted lines for electrons and ions, respectively.
While the electron spectra are fitted well by the Maxwellian distribution, the ion spectra exhibit a noticeable suprathermal shoulder that resembles a shifted Maxwellian, likely arising from pre-acceleration processes at the shock front~\citep{park_particle--cell_2012, caprioli2014SIMULATIONSIONACCELERATION}.  
Accordingly, based on the transition in particle spectra, we will universally use $\gamma_{\rm br,e} = 10\,T_e^{\rm d}/(m_i c^2)$ and $\gamma_{\rm br,i} = 10\,E_{\rm sh}/(m_i c^2)$ for all cases, with $T_e^{\rm d}$ denoting the electron downstream temperature.
The nonthermal tails follow a power-law distribution, $f(K) \propto K^{-s}$, with the spectral indices $s_i \approx s_e \simeq 2$, consistent with the predictions from DSA theory for strong shocks. In general, ion spectra produced by DSA are hard, with $s_i \lesssim 2$ in most cases and reaching values as low as $\sim 1.5$.
A power-law index $s = 1.5$ is expected in the nonrelativistic energy regime and gradually steepens toward $s_e \sim 2$ in the relativistic limit~\citep{haggerty2019DHybridRHybridParticleinCellCode}.  
In contrast, the electron spectral index varies significantly with shock parameters, becoming noticeably softer ($s \gtrsim 3$) for weaker shocks with $M_s \lesssim 8$ (e.g., runs S4T1, S4T10, and S4T50).  
Further details of the spectral properties across different simulations are summarized in Appendix~\ref{sec:Appendix_A}.

To illustrate the acceleration mechanism, we tracked representative electron and ion trajectories in the $x$--$\gamma$ space, as shown in panels (b) and (c) of Figure~\ref{fig:spec_track}.
Here, we normalized the particle positions relative to the shock front, whereas $\gamma$ is computed in the simulation frame.
The color of each trajectory corresponds to the simulation time and shares the same scale as the upper panel.
The ion trajectory shows that the particle resides predominantly in the upstream region and undergoes repeated scattering between the upstream and downstream regions. 
The ion Lorentz factor $\gamma$ increases each time the particle returns to the shock front after scattering off upstream fluctuations. 
This apparent ``energy gain'' occurs in the downstream frame; if transformed to the shock frame, the particle gains energy upon each shock crossing, as commonly described in DSA theory (see Figure~8 of \citealt{kato2015PARTICLEACCELERATIONWAVE}).

The electron's acceleration reveals a more complex, two-stage process.
The accelerated electron first undergoes a rapid pre-acceleration phase close to the shock front, with its trajectory resembling a hybrid shock drift acceleration (SDA), a process similar to that described in previous work~\citep{park2015SimultaneousAccelerationProtons}.
Once it attains sufficient energy from this initial stage, it is injected into the main DSA cycle for further energization in a manner similar to that of ions.
The significant mass difference between electrons and ions directly impacts their acceleration efficiency.
Due to their much smaller gyroradii, electrons require more DSA cycles to achieve the same energy as ions.
This lower efficiency results in a smaller nonthermal energy fraction for electrons compared to ions, as we can see from the energy spectra shown in Figure~\ref{fig:spec_track}(a).

\subsubsection{Streaming instabilities} \label{subsubsec:instability}

\begin{figure}[t]
     \centering
     \includegraphics[width=0.45\textwidth]{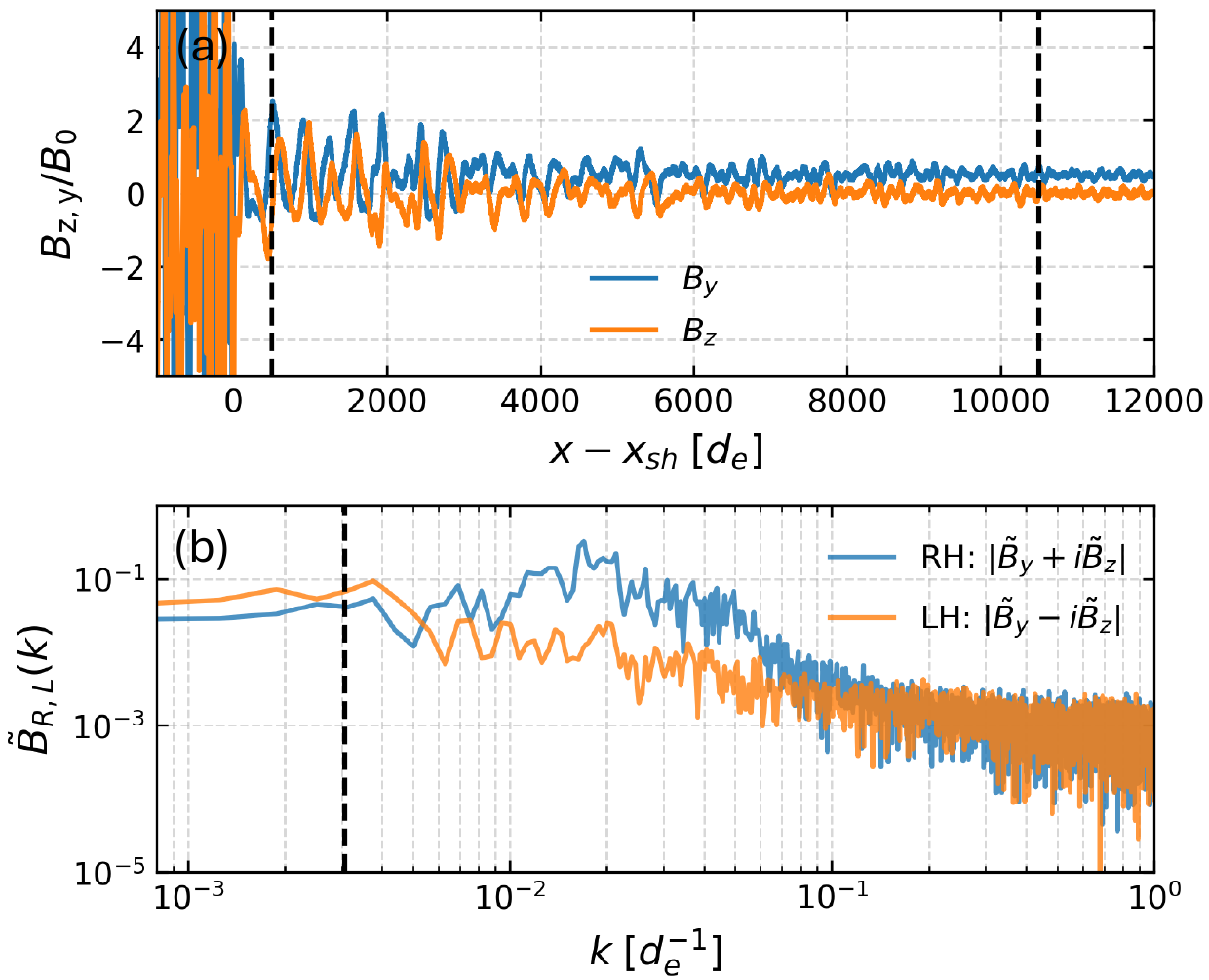}
     \caption{\label{fig:nres_instability}
     Spatial Fourier analysis of the upstream streaming instability at $t \approx 130~\Omega_{\rm ci}^{-1}$, showing that the dominant mode is the right-hand nonresonant instability.
     (a) Transverse magnetic field profile in the upstream region selected for the spectral analysis, indicated by the two vertical black dashed lines.
     (b) Fourier power spectrum of the magnetic field, decomposed into right-hand (RH) and left-hand (LH) polarized modes. 
     The vertical black dashed line marks the wavenumber of the mode resonant with backstreaming ions of mean gyroradius $\bar{\rho}_i$, $k_{\rm res} \sim 1/\bar{\rho}_i$.
     }
\end{figure}

The nature of the instabilities driven by streaming ions depends primarily on the ion current density~\citep{haggerty2019DHybridRHybridParticleinCellCode, marcowith2021CosmicRaydrivenStreaming}. 
In weak current regimes, the system is dominated by the resonant instability (RI), which is excited through a gyroresonant interaction between the streaming ions and Alfv\'enic fluctuations~\citep{skilling_cosmic_1971}. 
In contrast, at sufficiently high current densities, a strong neutralizing return current is carried by background electrons, and it excites modes at shorter wavelengths, leading to the nonresonant instability (NRI), also called Bell's instability~\citep{bell2004TurbulentAmplificationMagnetic}.
Kinetic plasma studies provide a practical method for distinguishing between them based on the wave polarization: the fastest-growing NRI modes are predominantly right-handed (RH), whereas for RI, the left-handed (LH) component is expected to be of greater magnitude~\citep{haggerty2019DHybridRHybridParticleinCellCode, gupta2024ElectronAccelerationQuasiparallel}.

To study the source of the upstream magnetic turbulence in run T10, we performed a spatial Fourier analysis of the transverse magnetic field in the upstream region at $t \approx 130~\Omega_{\rm ci}^{-1}$, as indicated by the two vertical black dashed lines in Figure~\ref{fig:nres_instability}(a). 
The corresponding magnetic amplitude spectra of the RH and LH polarized modes, defined respectively as $\tilde{B}_R = |\tilde{B}_y(k) + i\tilde{B}_z(k)|$ and $\tilde{B}_L = |\tilde{B}_y(k) - i\tilde{B}_z(k)|$, are shown in panel (b). 
Here $\tilde{B}_y,z$ is the spatial Fourier amplitude in the transverse directions.
The RH mode clearly dominates over the LH mode, exhibiting a significantly larger peak amplitude. 
The spectrum peaks at a wavenumber \(k_{\max}\) larger than the wavenumber of the mode resonant with backstreaming ions \(k_{\rm res} \sim 1/\bar{\rho}_i\) indicated by a vertical black dashed line in Figure~\ref{fig:nres_instability}(b).
Here, $\bar{\rho}_i$ is the mean gyroradius of the backstreaming ions, which can be estimated using $v\approx 5v_{\rm sh}/4$, corresponding to the velocity of reflected particles in the downstream frame of a strong shock.
This behavior is consistent with the expected characteristics of an NRI mode.

We performed a similar analysis for all simulation cases in the present study and found that almost all cases exhibit the NRI as the dominant mode. 
A few exceptions occur in the low-Alfvénic-Mach-number runs with $M_A = 4$ and $8$ (runs A4 and A8). 
This trend is consistent with previous numerical studies of quasi-parallel collisionless shocks, which report that low-$M_A$ shocks preferentially excite the RI~\citep{caprioli2014SIMULATIONSIONACCELERATIONa, gupta2024ElectronAccelerationQuasiparallel}. 
Nevertheless, both types of instabilities are capable of generating sufficient upstream magnetic turbulence to sustain DSA of both ions and electrons.

\subsection{Proton acceleration} \label{subsec:proton_acceleration}

We now extend our analysis to quantify the dependence of the proton energy gain or acceleration rate on various shock parameters.
In particular, we discuss how the acceleration timescale evolves and what implications it has for potential neutrino emission from the corona.

\begin{figure*}[t!]
\centering
\includegraphics[width=1\textwidth]{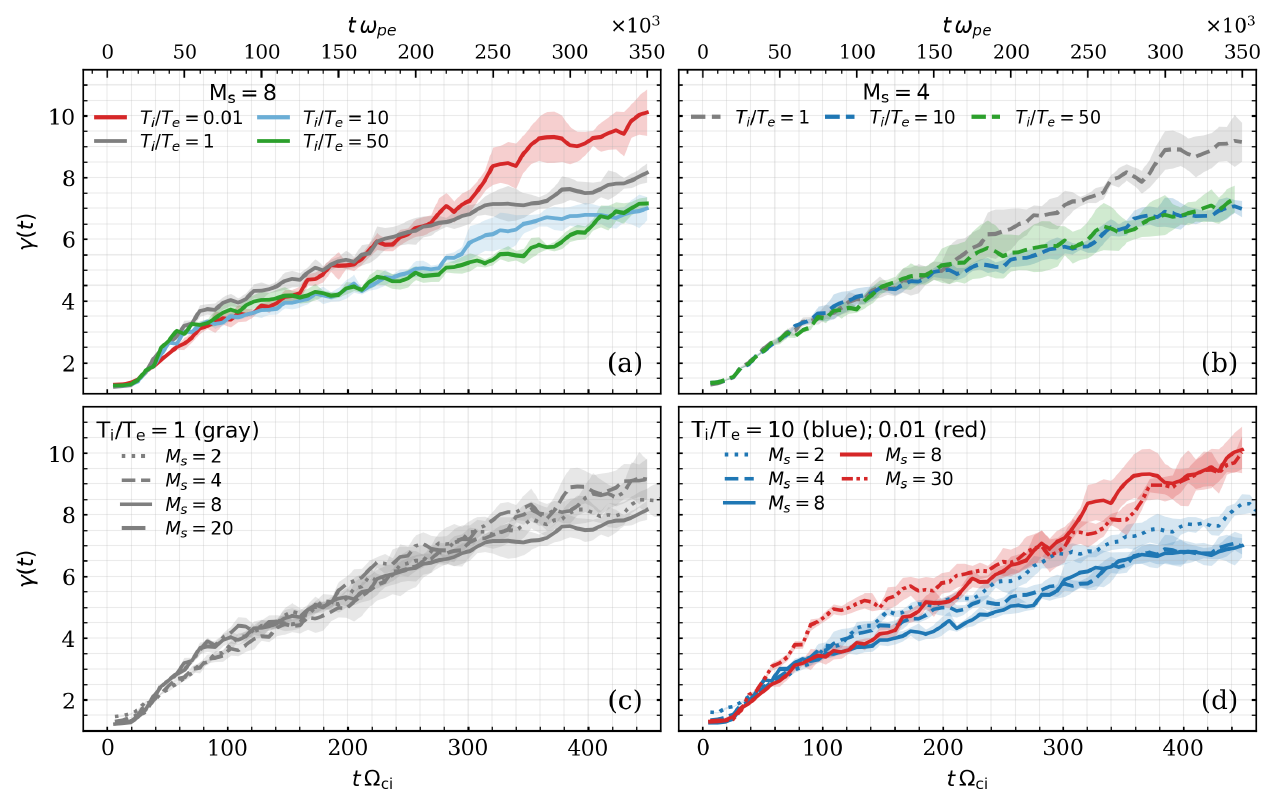}
\caption{
\label{fig:proton_gamma_t_TrMs}
Time evolution of the mean Lorentz factor $\gamma(t)$ of the 10 most energetic ions at each snapshot for different plasma conditions.
The upper panels show the effect of varying the temperature ratio $T_i/T_e$ at fixed $M_s$, while the lower panels show the effect of varying $M_s$ at fixed $T_i/T_e$.
Shaded regions represent one standard deviation.
All other parameters are identical to the reference run (T10), unless otherwise noted.
(a--b) Dependence on $T_i/T_e$ for $M_s = 8$ and $M_s = 4$, respectively.
(c) Dependence on $M_s$ for $T_i/T_e = 1$.
(d) Same as (c), but for both $T_i/T_e = 0.01$ (red lines) and $T_i/T_e = 10$ (blue lines).}
\end{figure*}

\subsubsection[Dependence on Ti/Te and Ms]{Dependence on ${T_i/T_e}$ and ${M_s}$}

To illustrate how the maximum ion energy evolves over time, Figure~\ref{fig:proton_gamma_t_TrMs} shows the time evolution of the average Lorentz factor $\gamma(t)$ of the 10 most energetic protons at each snapshot in each simulation.
All cases in Figure~\ref{fig:proton_gamma_t_TrMs} share the same initial drift velocity ($v_{\rm pt}/c = 0.25$) and magnetic field strength ($M_A = 26$).
The top panels display the simulations with different $T_i/T_e$ values at fixed Mach numbers: $M_s = 8$ in panel (a) and $M_s = 4$ in panel (b).
In contrast, the bottom panels compare cases with different $M_s$ values while keeping $T_i/T_e$ constant: $T_i/T_e = 1$ in panel (c), and $T_i/T_e = 10$ (blue lines) and $T_i/T_e = 0.01$ (red lines) in panel (d).

In all cases, the ion energy $\gamma(t)$ evolves in two distinct phases, separated around $t \sim 100~\Omega_{\rm ci}^{-1}$.
The early phase corresponds to the period required for the shock structure to reach a quasi-steady state and for upstream turbulence to fully develop.
In the later stage ($t \gtrsim 100~\Omega_{\rm ci}^{-1}$), the energy gain proceeds approximately linearly with time, $\gamma(t) \propto t$, consistent with the prediction of DSA theory \citep[e.g.][]{drury_introduction_1983, blandford_particle_1987}.
Figures~\ref{fig:proton_gamma_t_TrMs}(a) and (b) show that increasing the temperature ratio $T_i/T_e$ leads to a weak inverse correlation with the acceleration rate, roughly following $\gamma(t) \propto (T_i/T_e)^{-0.15}$.
However, this dependence is weak even between the extreme cases of $T_i/T_e = 0.01$ and $T_i/T_e = 50$. 
At the end of the runs, $\gamma(t)$ differs by a factor $\lesssim 1.5$.
The sonic Mach number $M_s$ also appears to have no effect on the maximum ion energy, as shown in Figures~\ref{fig:proton_gamma_t_TrMs}(c) and (d), where cases with the same $T_i/T_e$ (lines of the same color) converge to nearly identical values.

\subsubsection[Dependence on MA and vpt]{Dependence on $M_A$ and $v_{\rm pt}$}\label{subsubsec:M_A_v_pt}

\begin{figure}[t!]
\centering
\includegraphics[width=0.48\textwidth]{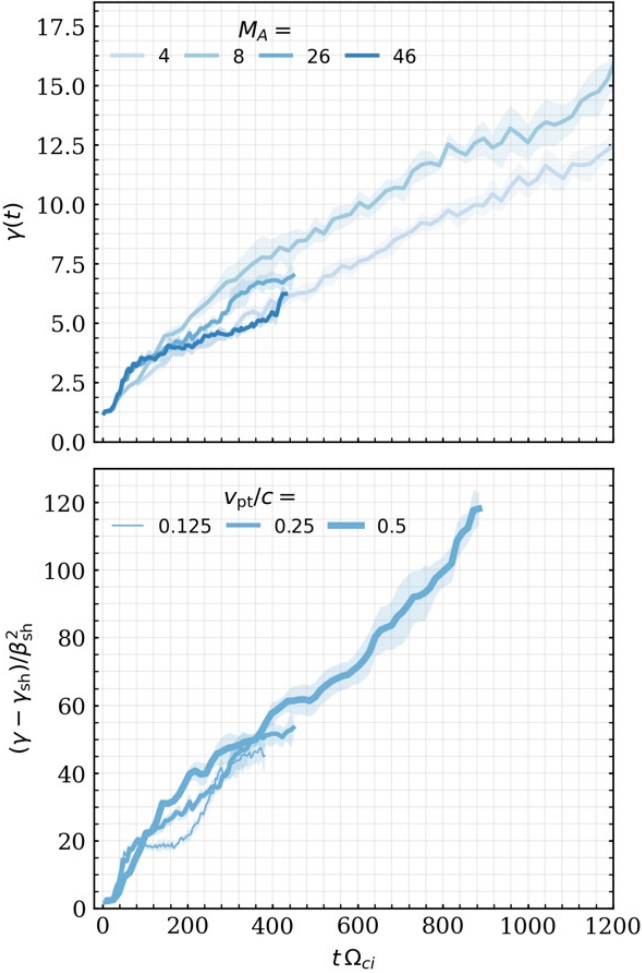}
\caption{
\label{fig:proton_gamma_t_VptMa}
Same format as Figure~\ref{fig:proton_gamma_t_TrMs}, but showing the dependence on the Alfvénic Mach number and the injected particle velocity. 
(a) Time evolution of the mean Lorentz factor of the 10 most energetic ions for different $M_A$. 
(b) Normalized energy gain, $(\gamma - \gamma_{\mathrm{sh}})\beta_{\mathrm{sh}}^{-2}$, for different injected velocities $v_{\rm pt}/c$. 
Shaded regions indicate one standard deviation. 
All other parameters are identical to the reference run (T10), unless otherwise specified.}
\end{figure}

We next examine the influence of the Alfvénic Mach number $M_A$ and the injected particle velocity $v_{\rm pt}$. 
According to DSA theory, the acceleration timescale depends on the magnetic field strength and the shock velocity as $t(\gamma) \simeq D(\gamma)/v_{\mathrm{sh}}^{2}$, where the diffusion coefficient is approximately Bohm-like, $D(\gamma) \simeq c E_{\rm CR} / (e B)$, where $E_{\rm CR} \simeq \gamma m_i c^2$
~\citep{drury_introduction_1983, blandford_particle_1987}. 
This yields the scaling
\begin{equation}
    t(\gamma) \approx \mathcal{C}\,\gamma\,\beta_{\mathrm{sh}}^{-2}\,\Omega_{\rm ci}^{-1},
    \label{eq:t_DSA_norm}
\end{equation}
where $\beta_{\mathrm{sh}} = v_{\mathrm{sh}}/c$ and $\mathcal{C}$ represents a dimensionless coefficient that encapsulates other factors governing the acceleration efficiency, such as the gyrofactor $\eta_g$ and acceleration efficiency $\eta_{\mathrm{acc}}$. 
Hence, we express time in units of $\Omega_{\rm ci}^{-1}$. 
When comparing cases with different shock velocities, we normalize the acceleration rate by $(\gamma-\gamma_{\mathrm{sh}})\beta_{\mathrm{sh}}^{-2}$ in order to factor out the expected DSA scaling [Figure~\ref{fig:proton_gamma_t_VptMa}(b)].

Figures~\ref{fig:proton_gamma_t_VptMa}(a) and (b) illustrate the effects of varying $M_A$ and $v_{\rm pt}/c$, respectively.
As shown in Figure~\ref{fig:proton_gamma_t_VptMa}(a), cases with different Alfvénic Mach numbers exhibit broadly similar behaviors, except for the very low $M_A = 4$ case, which shows a slightly reduced acceleration rate compared to the reference run ($M_A = 26$).
This is likely due to weaker self-generated turbulence, as confirmed by the presence of resonant streaming instability~\citep{gupta2024ElectronAccelerationQuasiparallel}.
Such a trend is expected, as the shock's Alfvénic Mach number must exceed order unity for efficient particle acceleration to occur (see \cite{vink_critical_2013}, who predicted a threshold of $M_A \simeq 5/2$).
Accordingly, the acceleration rate, ${\cal{C}}^{-1}$, is expected to decrease significantly as $M_A \rightarrow 1$.
When the injection velocity varies from $v_{\rm pt}/c = 0.125$ to $0.5$ in  Figure~\ref{fig:proton_gamma_t_VptMa}(b), the normalized acceleration rates in all three runs remain remarkably similar, confirming that the acceleration efficiency, ${\cal{C}}$, follows the scaling predicted by Equation~(\ref{eq:t_DSA_norm}).

\begin{figure*}[t!]
\centering
\includegraphics[width=0.98\textwidth]{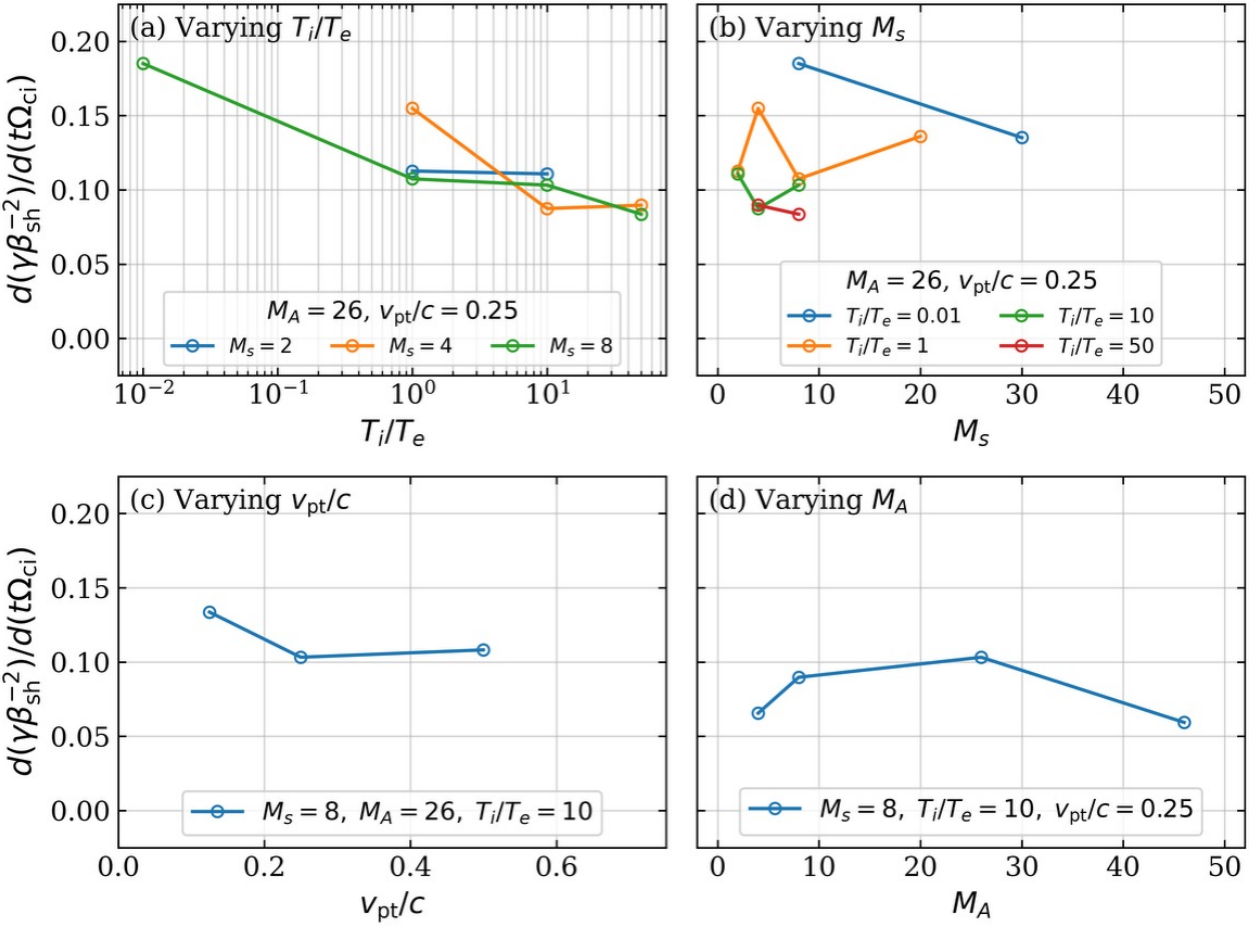}
\caption{\label{fig:acc_coeff} The normalized acceleration coefficient $d(\gamma/\beta_{\rm sh}^2)/d(t\Omega_{\rm ci})$ or the slope in Figures~\ref{fig:proton_gamma_t_TrMs} and~\ref{fig:proton_gamma_t_VptMa}, as a function of key shock parameters, calculated from the linear slope of the energy gain at late times. (a) Dependence on $T_i/T_e$. (b) and (c) Dependence on $M_s$ and $M_A$, respectively. (d) Dependence on $v_{\rm pt}/c$.}
\end{figure*}

\subsubsection{Scaling of acceleration rate}

We next compare the normalized acceleration rate,
$\mathcal{C}^{-1} = d(\gamma\beta_{\rm sh}^{-2})/d(t\Omega_{\rm ci})$, 
shown in Figure~\ref{fig:acc_coeff}. 
This coefficient is obtained from the slope of the mean energy gain during the late stage of acceleration [$t \gtrsim 100~\Omega_{\rm ci}^{-1}$; see, e.g., Figure~\ref{fig:proton_gamma_t_VptMa}(b)]. 
As clearly seen in Figure~\ref{fig:acc_coeff}, the typical value of the normalized acceleration rate is $\mathcal{C}^{-1} \simeq 0.1$, varying by less than a factor of two across all parameter sets.

The reference parameters for Figure~\ref{fig:acc_coeff} are $M_A=26$, $M_s=8$, $T_i/T_e=10$, and $v_{\rm pt}/c=0.25$. 
Each panel varies one parameter while keeping the others fixed: $T_i/T_e$ in (a) (colored lines for different $M_s$), $M_s$ in (b) (colored lines for different $T_i/T_e$), $v_{\rm pt}/c$ in (c), and $M_A$ in (d). 
The results show no clear dependence on $M_s$ [Figure~\ref{fig:acc_coeff}(b)] and only a weak, nearly flat trend with $v_{\rm pt}/c$ [Figure~\ref{fig:acc_coeff}(c)]. 
A slight systematic decrease is observed as ions become hotter relative to electrons (i.e., with increasing $T_i/T_e$). 

Panel (d) indicates a modest dependence of the acceleration rate on the Alfvénic Mach number. While the acceleration efficiency is somewhat reduced at both low and high $M_A$, it remains comparable over the full range considered.
The reduced acceleration efficiency at low $M_A$ likely results from weaker self-generated turbulence, as discussed in Section~\ref{subsubsec:M_A_v_pt}.
At very high $M_A$, however, the strong turbulence can excite short large-amplitude magnetic structures, which intermittently render parts of the shock superluminal and disrupt ion acceleration (see~\cite{zekovic_slams-propelled_2025} for a detailed discussion). 
This trend suggests the existence of an optimal magnetic field strength for efficient ion acceleration by DSA in this regime.

To summarize, we have demonstrated that ions are efficiently accelerated over a broad range of physical conditions, indicating a robust capability of shocks for ion energization in BH coronae.
Remarkably, the normalized acceleration rate varies by less than a factor of two across all tested parameters, consistent with the expectations from DSA theory.

\subsection{Energy partition} \label{subsec:energy_partition}

\begin{figure*}[t]
     \centering
     \includegraphics[width=0.99\textwidth]{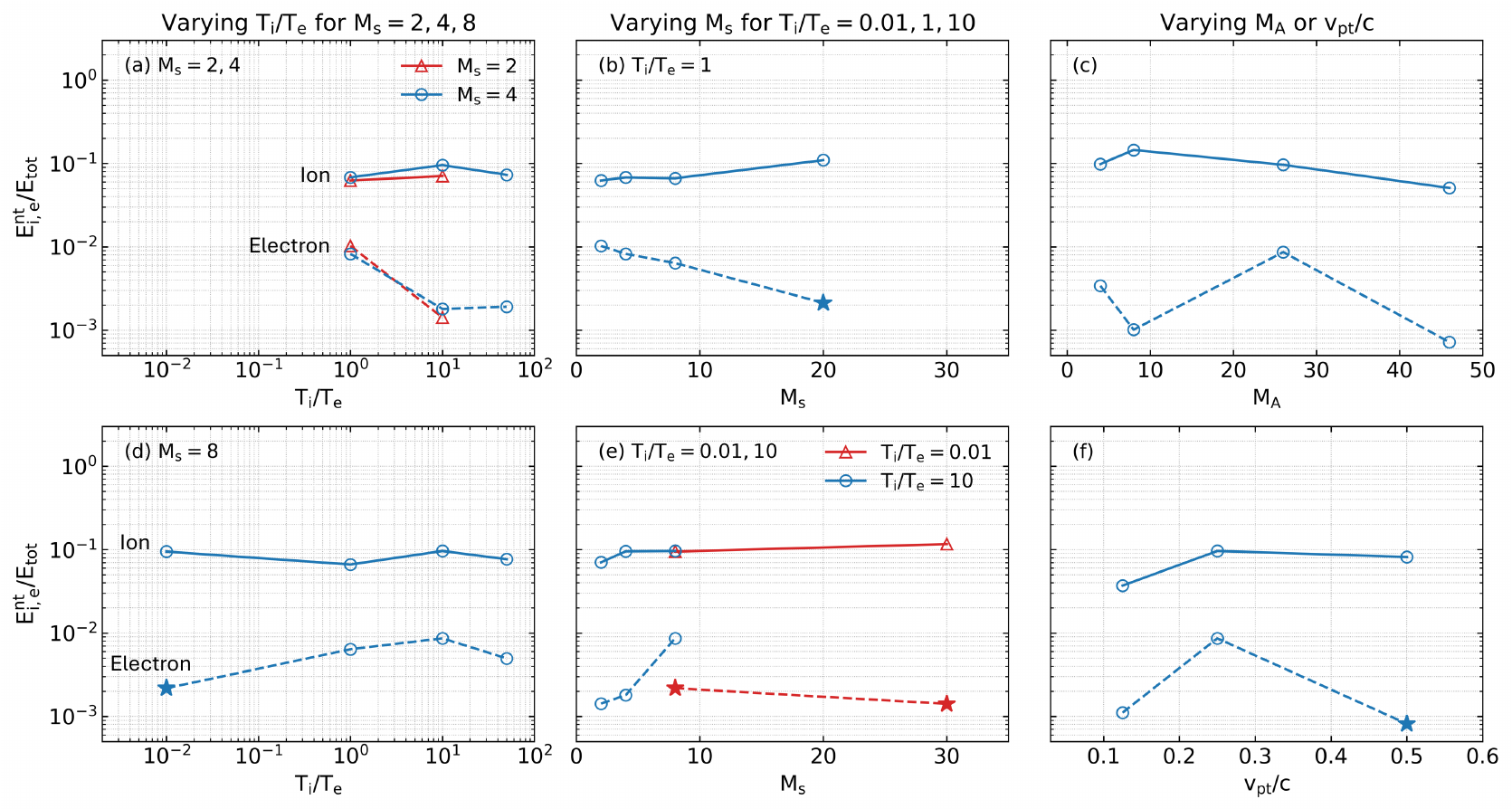}
\caption{\label{fig:energy_partition}Fraction of nonthermal ion (solid lines) and electron (dashed lines) energy to total energy in the particle spectra across the parameter space. Nonthermal ions are efficiently accelerated, receiving $\sim 10\%$ of the total energy, while electrons remain much less efficient ($<1\%$) or even suppressed in cases indicated by the $\star$ symbol. Left panels: the effect of the $T_i/T_e$ ratio with fixed $M_s = 2$ (red lines) and $M_s = 4$ (blue lines) in panel (a), and $M_s = 4$ in panel (d). Middle panels: the scaling with $M_s$ when $T_i/T_e$ is fixed at $T_i/T_e = 1$ in panel (b), and at $T_i/T_e = 0.01$ (red) and $T_i/T_e = 10$ in panel (e). Right panels: variation with (c) $M_A$ and (f) $v_{\rm pt}/c$. Unless explicitly stated otherwise, other parameters follow $T_i/T_e=10$, $M_s = 8$, $M_A = 26$, and $v_{\rm pt}/c = 0.25$.}
\end{figure*}

In this section, we investigate the partitioning of the shock energy into nonthermal ions and electrons (i.e., the acceleration efficiency), which is crucial for determining the relative power available for hadronic neutrino production and leptonic emission. 

We define the energy fraction, or the acceleration efficiency, as
\begin{equation}
    \frac{E^k_j}{E_{\rm tot}} = \frac{1}{E_{\rm tot}} \int d\gamma (\gamma - 1)m_jc^2f^k_j(\gamma),
    \label{eq:energy_fraction}
\end{equation}
where $j\in \{i,e\}$ and $k \in \{ \rm nt, th\}$ denote four different particle energy fractions $E^{\rm nt}_i, E^{\rm nt}_e,E^{\rm th}_i~\text{or}~E^{\rm th}_e$, corresponding to energy fractions of nonthermal ions, nonthermal electrons, thermal ions, and thermal electrons, respectively.
The total particle energy is given by $E_{\rm tot} = \sum^k_j \int d\gamma\, (\gamma - 1)m_jc^2f^k_j(\gamma)$, i.e., the sum of all particle energy.
We used the same distinction between thermal and nonthermal populations of ions and electrons as in Section~\ref{subsubsec:spectrum_trajectory}.

The detailed scaling of the energy fractions of nonthermal ions (solid lines) and electrons (dashed lines) is depicted in Figure~\ref{fig:energy_partition}.
Each line in the panels then represents a parameter scan, varying one of these values while all others are held constant.
Unless otherwise noted, the default parameters are those of the fiducial run T10: $M_A = 26$, $M_s = 8$, $v_{\rm pt}/c = 0.25$, and $T_i/T_e=10$.
We present results for the scaling with varying $T_i/T_e$ (also with a few $M_s$ values) in the left panels (a) and (d); with varying $M_s$ (with three $T_i/T_e$ values) in the middle panels (b) and (e); and with $M_A$ and $v_{\rm pt}/c$ in panels (c) and (f), respectively.

Remarkably, we find that nonthermal ions are efficiently accelerated, consistently receiving $\sim 10\%$ of the total energy.
This high ion acceleration efficiency remains robust across all parameter surveys, even in the low sonic Mach number regime where inefficient acceleration is often assumed.
As shown in the left (a, d) and middle (b, e) panels, the ion energy fraction $E^{\rm nt}_i/E_{\rm tot}$ remains nearly independent of $T_i/T_e$ and $M_s$ across all tested regimes.
Similarly, Figures~\ref{fig:energy_partition} (c) and (f) show a weak dependence on $M_A$ and $v_{\rm pt}/c$, with the ion acceleration efficiency slightly reduced at high $M_A$ [panel (c)] and for slower shocks [panel (f)].

The electron acceleration efficiency is much lower than that of ions, with $E^{\rm nt}_e/E_{\rm tot}$ remaining below $1\%$.
It also exhibits a more complex dependence, being highly sensitive to the upstream plasma parameters.
A major contributing factor is that, in some cases with $v_{\rm pt}/c \ge 0.25$, we found the maximum energy of the electron power-law tail does not continue to grow with time, leading to a significant reduction in the acceleration efficiency.
We mark such cases with the star symbol in Figure~\ref{fig:energy_partition}, for example, in the $T_i/T_e = 0.01$ runs in panel (e) (red dashed line) and the $v_{\rm pt}/c = 0.5$ run in panel (f).
The physical origin of this behavior is briefly discussed in Section~\ref{subsection:disscus_electron}; however, because electrons remain energetically subdominant, this effect does not affect the ion acceleration results or the overall energy partition discussed above.

Most notably, for implications relevant to BH coronae, $E^{\rm nt}_e/E_{\rm tot}$ shows a monotonically decreasing trend with increasing ion-to-electron temperature ratio $T_i/T_e$, as seen in panels (a) and (d).
For $T_i/T_e = 1$, the efficiency is $\sim 10^{-2}$, whereas for high $T_i/T_e \gtrsim 10$, it decreases to $\sim 10^{-3}$.
This result agrees with recent work on the electron injection threshold in quasi-parallel shocks~\citep{gupta2025SpeeddependentThresholdElectron}, which indicates that higher electron-thermal speeds in the immediate upstream region favor more efficient electron acceleration.
The decline of electron acceleration in high $T_i/T_e$ environments, which is expected in BH coronae, imposes a strong constraint on the multi-messenger emission from these regions.
In particular, it sets the limits on the energy budget for neutrino production by the hadronic component, as well as the primary leptons' budget for the synchrotron and inverse-Compton emissions.

\section{Discussion} \label{sec:discussion}

In this section, we first synthesize our results on proton acceleration in order to examine the case of NGC~1068, the most prominent candidate in the IceCube neutrino survey. We then briefly discuss electron acceleration, or rather the lack thereof, in our simulations.

\subsection{Implication for neutrino emissions from NGC~1068} \label{subsection:NGC1068}

Here, we explore the implications of our results for neutrino emissions from NGC~1068---the most prominent detection from the IceCube observatory to date~\citep{icecubecollaboration*+2022EvidenceNeutrinoEmission}.
The detection is constrained to be in the neutrino energy range $E_\nu = 1.5 - 15~\text{TeV}$, which implies the presence of protons with an energy range $E\approx 30-300~\text{TeV}$.
To reach such energies, the proton acceleration timescale must be shorter than the timescales associated with energy losses from cooling and neutrino production.
We therefore begin by comparing these relevant timescales.
Then, we examine whether the acceleration efficiency of protons inferred from our results (Section~\ref{subsec:energy_partition}) can fulfill the energy budget for the neutrino spectrum from the object.

In the following analysis, we adopt a set of coronal parameters for NGC~1068 listed in \cite{padovani2024HighenergyNeutrinosVicinity}, which we use consistently throughout this discussion unless stated otherwise. 
We assume a black hole mass of $M_{\rm BH} = 10^{7.2}\,M_\odot$ and a characteristic coronal size of $R_c = 20\,R_g$.
The coronal photon field consists of a multi-temperature blackbody emission from the accretion disk \citep{shakura1973BlackHolesBinary}, parameterized as
$L_{\rm disk} \propto E^{4/3} \exp(-E/E_{\rm cut})$ with $E_{\rm cut} = 30~\mathrm{eV}$,
together with a Comptonized X-ray component described by
$L_X \propto E^{1-\Gamma_X} \exp(-E/E_{\rm cut,X})$,
where $E_{\rm cut,X} = 128~\mathrm{keV}$ and $\Gamma_X = 1.95$.
Assuming a luminosity distance of $d_L = 10~\mathrm{Mpc}$, this corresponds to an integrated disk luminosity of $L_{\rm disk} = 5 \times 10^{44}~\mathrm{erg\,s^{-1}}$ and a $2$--$10~\mathrm{keV}$ X-ray luminosity of $L_{2-10\,\mathrm{keV}} = 4 \times 10^{43}~\mathrm{erg\,s^{-1}}$.

\subsubsection{Timescales}
\begin{figure}
    \centering
    \includegraphics[width=0.45\textwidth]{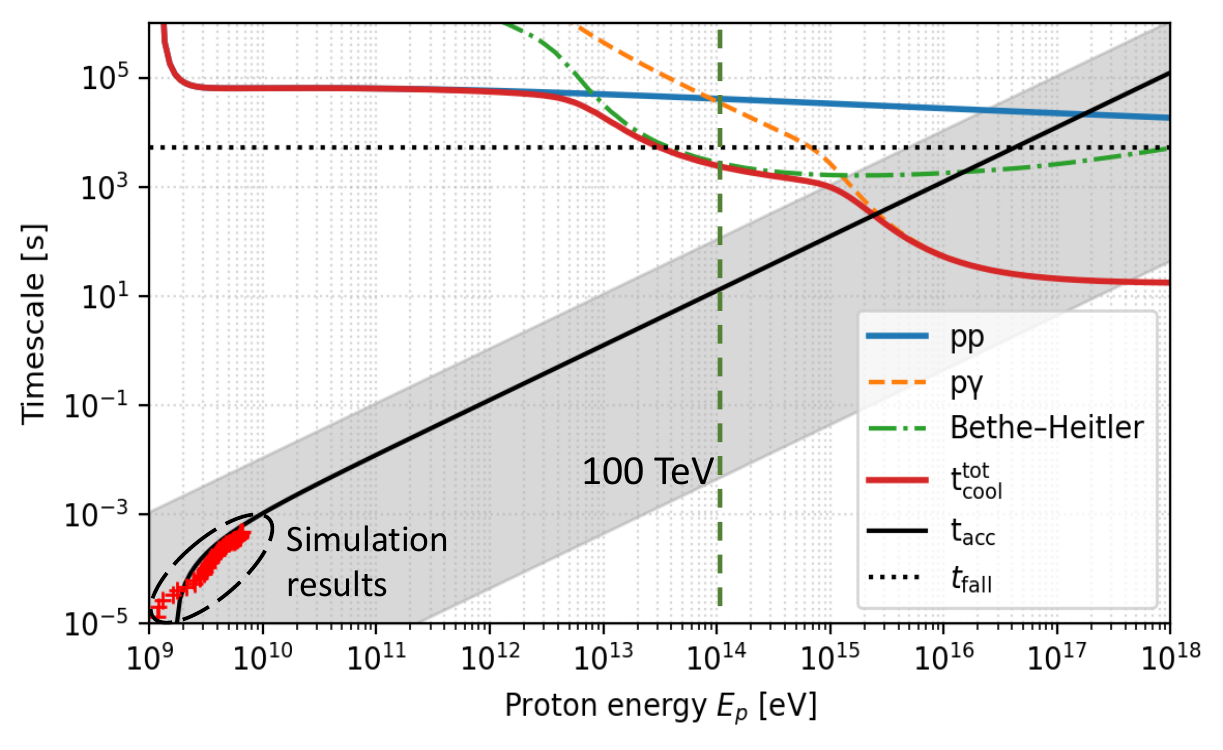}
    \caption{\label{fig:timescales_NGC1068} Various timescales in the BH corona of NGC~1068 as a function of proton energy.
    The typical acceleration timescale $t_{\rm acc}$ (solid black line) is calculated using Equation~(\ref{eq:t_acc}) with $B_0 = 100~\text{G}$ and $\beta_{\rm sh} = 0.32$.
    The free-fall timescale $t_{\rm fall}$ (dotted black line) is calculated using Equation~(\ref{eq:t_free_fall}).
    The simulation results taken from run T10 give the normalization of $t_{\rm acc}$.
    The gray shade indicates the acceleration timescale range when we vary $B_0 = 10 - 1000~\text{G}$ and $\beta_{\rm sh} = 0.1 - 0.5$.
    The cooling timescales for $pp$ interactions $t_{pp}$ (blue line), $p\gamma$ interactions $t_{p\gamma}$ (orange dashed line), and Bethe-Heitler pair production $t_{\rm BH}$ (green dashed-dotted line) are calculated following the methods in~\cite{inoue2019HighenergyParticlesAccretion}.
    The total cooling timescale $t_{\rm cool}$ (red line) is obtained by combining all cooling processes.
    We include the vertical green dashed line showing protons of 100 TeV energy, which are required to explain TeV neutrinos.}
\end{figure}

In the scenario shown in Figure~\ref{fig:schematic} discussed in Section~\ref{sec:condition_BHCS}, the dynamical timescale is determined by the free-fall time,

\begin{equation}
    t_{\rm fall} = \frac{R_c}{v_{\rm ff}} \simeq 0.52\times10^4 
    \left( \frac{r_c}{20} \right)^{3/2} 
    \left( \frac{M_{\rm BH}/M_{\odot}}{10^{7.2}} \right)\;[\text{s}]\;.
    \label{eq:t_free_fall}
\end{equation}
All the related timescales are summarized in Figure~\ref{fig:timescales_NGC1068} as a function of proton energy.

To account for the observed neutrino emission, protons must be accelerated to energies of $E_p \sim 100~\mathrm{TeV}$. 
At these energies, in addition to hadronic pion production via $pp$ and $p\gamma$ interactions, protons also experience energy losses through the Bethe--Heitler pair production process.
We compute the corresponding cooling timescales for Bethe--Heitler ($t_{\rm BH}$), photomeson ($t_{p\gamma}$), and proton--proton ($t_{pp}$) interactions using the coronal photon field and parameters defined above.
The detailed calculation procedures follow those outlined in \citet{inoue2019HighenergyParticlesAccretion}.

For proton acceleration timescale $t_{\rm acc}$, we make use of our result obtained from Section~\ref{subsec:proton_acceleration}.
In particular, we adopt the scaling of Equation~(\ref{eq:t_DSA_norm})
with the typical value of $\mathcal{C}^{-1}=d(\gamma\beta_{\rm sh}^{-2})/d(t\Omega_{\rm ci}) \simeq 0.1$.
The scaling for proton acceleration timescale reads
\begin{equation}
    \begin{split}
    t_{\rm acc} & \simeq \mathcal{C}  \gamma \beta_{\rm sh}^{-2} \Omega_{\rm ci}^{-1} \\
                & \simeq 11.1
                \left(\frac{E_p}{100~\text{TeV}} \right)
                \left(\frac{\mathcal{C}^{-1}}{0.1}\right)^{-1}
                \left(\frac{\beta_{\rm sh}}{0.32} \right)^{-2}
                \left(\frac{B_0}{100~\text{G}} \right)^{-1}
                \;[\text{s}]\;.
    \end{split}
    \label{eq:t_acc}
\end{equation}
In Figure~\ref{fig:timescales_NGC1068}, we extrapolate the simulation scaling given by Equation (\ref{eq:t_acc}) to several orders of magnitude, assuming continuous evolution.

In addition to the balance between acceleration and cooling timescales, the maximum proton energy is constrained by the finite size of the BH corona region. This geometrical constraint, $\rho_i \lesssim R_c$, is known as the Hillas limit~\citep{hillas_origin_1984} and implies a maximum achievable energy $E_{\rm max} \lesssim e B_0 R_c$.
In the context of DSA, a tighter constraint is often applied, taking the form $E_{\rm max} \lesssim e B_0 R_c \beta_{\rm sh}$~\citep{drury2012OriginCosmicRays, oka2025MaximumEnergyParticles}, which yields the following scaling
\begin{equation}
    E_{p,{\rm max}} \lesssim 467 \left(\frac{B}{100~\text{G}} \right)
                                \left(\frac{r_c}{20} \right)
                                \left(\frac{M_{\rm BH}/M_{\odot}}{10^{7.2}} \right)
                                \left(\frac{\beta_{\rm sh}}{0.32} \right)\;[\text{PeV}]\;.
    \label{eq:E_p_Hillas}
\end{equation}
Equation~(\ref{eq:E_p_Hillas}) shows that the Hillas criterion allows proton acceleration up to energies of $\sim100$ TeV, provided that the magnetic field strength satisfies $B_0 \gtrsim 10$ G and the shock velocity fulfills $\beta_{\rm sh} \gtrsim 10^{-2}$. 
Moreover, as shown in Figure~\ref{fig:timescales_NGC1068}, the acceleration timescale $t_{\rm acc}$ remains shorter than the total cooling timescale $t_{\rm cool}$ up to proton energies of $E_p \sim 1\,\mathrm{PeV}$ for the same range of parameters. 
Hence, these results demonstrate that our DSA framework can provide a robust mechanism for supplying the high-energy protons responsible for the high-energy neutrinos in the energy range observed by IceCube.

We note that this simple extrapolation would yield maximum energies higher than those inferred from IceCube observations. However, extrapolation over several orders of magnitude is inherently uncertain, and a more detailed investigation of the maximum energy is required.

\subsubsection{Energy budget}
We will now discuss whether the acceleration efficiency derived from our simulations ($E^{\rm nt}_p/E_{\rm tot} \gtrsim 10\%$) is sufficient to produce the observed neutrino flux.
In the scenario of a spherical shock with $R_{\rm sh} \simeq R_c$, as we assume here, the bound of shock power can be simply estimated as
\begin{equation}
    \begin{split}
    L_{\rm sh} &= \frac{1}{2} n_i m_i v_{\rm sh}^3 (4\pi R_{\rm sh}^2) \\
               &\simeq 7.3 \times 10^{44} 
                \left(\frac{r_c}{20} \right)
                \left(\frac{M_{\rm BH}/M_{\odot}}{10^{7.2}} \right)
                \left(\frac{v_{\rm sh}/c}{0.32} \right)^3 ~[\text{erg/s}]\;,
    \label{eq:L_sh}
    \end{split}
\end{equation}
where we also assume $n_i \simeq n_e \simeq (\sigma_T R_c)^{-1}$.

According to the latest IceCube report~\citep{abbasi_evidence_2025}, the best-fit flux at 1 TeV is $\Phi_{\nu} = 4.7^{+1.1}_{-1.3} \times 10^{-11}~\text{TeV}^{-1}~\text{cm}^{-2}~\text{s}^{-1}$, with a neutrino luminosity $L_{\nu} \approx 4\pi d_L^2 E_{\nu}^2 \Phi_{\nu} \simeq 9 \times 10^{41} \times [d_L/(10~\text{Mpc})]^2~\text{erg/s}$.
The nonthermal proton luminosity required to produce such neutrino luminosity can be estimated as
\begin{equation}
    L_p^{\rm nt} \approx \frac{\mathcal{S} L_{\nu}}{f_{\nu}} \;,
    \label{eq:L_p_nt}
\end{equation}
where $f_{\nu}$ is the per flavor production fraction, and $\mathcal{S} \geq 1$ is a spectral-shape factor that accounts for the continuous nature of the nonthermal proton spectrum.
Neutrinos with $E_\nu = 1 ~\text{TeV}$ are produced by protons with energies $E_p \sim 20 E_\nu \sim 20~\text{TeV}$, which mainly produce neutrinos through $pp$ interactions (Figure~\ref{fig:timescales_NGC1068}).
Hence, we adopt $f_{\nu} \sim 1/6$ and $\mathcal{S} \gtrsim 4$ for a proton spectrum with power-law index $s_p = 2$ as seen in most of our simulations.
The required nonthermal proton luminosity is therefore
\( L_p^{\rm nt} \approx 2.2 \times 10^{43}~\mathrm{erg\,s^{-1}} \), corresponding to a fractional power
\( L_p^{\rm nt}/L_{\rm sh} \approx 0.03 \),
which is below the nonthermal ion energy fraction of \(\sim 10\%\) measured in our simulations at late times.
We emphasize that this \(\sim 10\%\) is obtained directly from the downstream particle energy distribution within the resolved energy range of the simulations at a given time, and is not derived by extrapolating the proton spectrum to the highest energies.
With this caveat, our estimate supports the possibility that collisionless shocks may contribute significantly to the proton power required for high-energy neutrino production in AGN coronae.

Finally, we want to stress that this estimation aims to put our results into perspective and provide an intuitive picture of neutrino production in NGC~1068.
We acknowledge that applying results from local, first-principles kinetic simulations to a complex astrophysical object such as an AGN corona requires careful consideration.
In this work, we have self-consistently determined the physical conditions at the localized shock front from first principles.
The estimation presented here should be regarded as an order-of-magnitude assessment based on the microphysics established in the current paper.
Nonetheless, it offers valuable insights into how the local shock acceleration process can contribute to the global energetics of the source and bridge the gap between microscopic plasma physics and macroscopic high-energy phenomena observed by IceCube.

\subsection{On electron acceleration} \label{subsection:disscus_electron}

We will next briefly comment on the simulations marked with a star symbol in Figure~\ref{fig:energy_partition}, in which the nonthermal electron power-law tail does not extend to progressively higher energies with time, as noted in Section~\ref{subsec:energy_partition}. 
In the trans-relativistic shock regime explored in this work, strongly amplified magnetic fields produced by ion-streaming instabilities can make the shock intermittently superluminal.
Electrons with gyroradii smaller than the saturated wavelength of the ion-streaming instabilities can be affected by the amplification \citep{jikei_magnetic_2025}.
In such cases, electrons are unable to propagate far upstream of the shock and repeatedly cross it, leading to a suppression of DSA.
Then, the maximum energy will be set by pre-acceleration mechanisms operating close to the shock front, such as SDA, rather than growing with time.

Our results are consistent with previous findings for high-shock-velocity cases \citep{gupta2024ElectronAccelerationQuasiparallel} and for shocks embedded in strongly amplified, locally superluminal magnetic fields \citep{zekovic_slams-propelled_2025, jikei_magnetic_2025}. 
In these regimes, pre-acceleration processes operating near the shock can still generate nonthermal electron power-law tails extending up to $\gamma_e\sim10^2$.
However, in the absence of sustained DSA, the maximum electron energy saturates and does not increase secularly with time, being limited by the characteristic scale of the pre-acceleration mechanisms or by the typical wavelength of the amplified magnetic fluctuations.
We further note that this effect is not expected to have a strong impact on black hole coronae, as such suppression occurs only for shock parameters associated with extremely strong magnetic field amplification ($\delta B / B_0 \gg 1$), including low ion temperatures (i.e., low $M_s$ or $T_i/T_e$; see also Figure~\ref{fig:energy_partition}) or very large Alfv\'enic Mach numbers ($M_A \gg 1$).
These conditions are unlikely to be realized in BH coronal environments.

Finally, the generally low electron acceleration efficiency at large $T_i/T_e$ can be attributed to the injection conditions for electrons at quasi-parallel shocks.
Previous studies have pointed out that the electron-thermal speed sets the threshold for injection into acceleration processes~\citep{gupta2025SpeeddependentThresholdElectron}, and that initially cold electrons must be pre-heated by another mechanism before they can be injected~\citep{gupta2024ReturnCurrentsCollisionless}.
Therefore, we naturally expect that a higher $T_e$ (i.e., lower $T_i/T_e$) would be more favorable for electron acceleration.
It remains unclear whether a much longer simulation would show that preheating of background electrons in the upstream region could increase their temperature sufficiently near the shock for efficient acceleration.
A detailed investigation of electron suppression by strong amplified field, preheating, and acceleration signatures in these regimes is beyond the scope of the present paper and will be addressed in our upcoming work.

\subsection{Comparison with previous PIC studies}

Our kinetic survey indicates that trans-relativistic collisionless shocks with a low sonic Mach number in a hot coronal plasma can efficiently inject roughly $10\%$ of the total shock energy into nonthermal protons for neutrino production. This result is broadly consistent with recent findings on quasi-parallel trans-relativistic shocks~\citep{pcrumley2019KineticSimulationsMildly, gupta2025SpeeddependentThresholdElectron, jikei_magnetic_2025}, although differences in the inferred ion acceleration efficiency may arise from, for example, the definition of the threshold separating thermal and nonthermal populations (i.e., the break Lorentz factor $\gamma_{\rm br}$).

Building on these previous studies, our work examines particle acceleration in the hot plasma environments of BH coronae, which are characterized by much lower sonic Mach numbers than those typically considered in earlier works. In particular, we extend the explored parameter space down to trans-sonic shocks with $M_s = 2$. Our goal is to determine whether shocks in this regime, often described in the literature as less efficient at accelerating particles~\citep{ha_proton_2018, kang_electron_2019, vanmarle_influence_2020}, can still serve as a viable mechanism for proton acceleration and the subsequent neutrino production observed by IceCube. In addition, we systematically explore the dependence on the upstream temperature ratio $T_i/T_e$. Since ion and electron heating differ in BH coronae~\citep{Kawazura2019, gorbunov2024FirstprinciplesMeasurementIon}, and the ion temperature $T_i$ remains uncertain due to the lack of direct observational constraints, a systematic survey of $T_i/T_e$ is both important and has not been explored in prior PIC studies.

Regarding the nature of upstream instabilities, \citealt{jikei_magnetic_2025} suggest that the Weibel instability may dominate over the NRI when the shock magnetization is $\sigma \lesssim 10^{-4}$, where $\sigma = B_0^2 / [4\pi (m_i + m_e)c^2]$. In our model, the magnetization parameter is on the order of $\sim 10^{-3} - 10^{-4}$. Because our survey is based on 1D3V simulations, it does not capture the transverse filamentation physics driving the Weibel instability, and thus cannot fully address its possible role during the earliest stage of shock formation. However, this limitation does not necessarily undermine the relevance of our model to later stages of evolution. Previous kinetic studies (e.g., \citealt{pcrumley2019KineticSimulationsMildly}) have shown that shocks may initially be mediated by the Weibel instability before gradually transitioning to a Bell-mediated (NRI) regime that regulates steady-state particle acceleration. Furthermore, simulations with $\sigma = 10^{-3}$ (corresponding to $M_A \simeq 26$) in \citealt{jikei_magnetic_2025} also indicate that the NRI eventually becomes the dominant mode. The interplay between these two instability modes, and whether such a transition persists at the lower shock velocities expected in BH coronae, compared to the faster shocks ($v_{\rm sh} / c \gtrsim 0.8$) explored in \citealt{pcrumley2019KineticSimulationsMildly} and \citealt{jikei_magnetic_2025}, remains uncertain. Nevertheless, we anticipate that the NRI captured in our simulations may still be relevant for determining the late-stage particle acceleration.

\section{Conclusion} \label{sec:conclusion}

Recent IceCube observations of TeV neutrinos from the Seyfert galaxy NGC~1068 suggest efficient proton acceleration within the BH coronae. To investigate the viability of DSA in this context, we systematically performed a series of 1D3V PIC simulations of quasi-parallel shocks. Our parameter survey covers the specific regimes motivated by BH coronae, including trans-relativistic shock velocities and varying ion-to-electron temperature ratios ($T_i/T_e$), while systematically varying the sonic ($M_s$) and Alfvénic Mach numbers ($M_A$).

Our primary findings are as follows:

\begin{enumerate}
    \item \textbf{Strong Shocks and Turbulence Generation:} Under coronal conditions, strong shocks can form and drive intense upstream magnetic fluctuations mediated by backstreaming ions, providing the scattering environment necessary for DSA. The turbulence is predominantly driven by the NRI across most of the parameter space, with RI modes becoming significant only at the lowest Alfvénic Mach numbers.
    
    \item \textbf{Proton Acceleration Dynamics:} We identify clear nonthermal ion power-law tails where the maximum energy increases linearly with time after the establishment of a quasi-steady shock structure. The proton acceleration rate scales with the square of the shock velocity, $(v_{\rm sh}/c)^2$, and linearly with the magnetic field strength (via the proton gyrofrequency, $\Omega_{\rm ci}$). This scaling is consistent with the standard Bohm-type diffusion in DSA theory.
    
    \item \textbf{Universal Acceleration Rate:} Based on this scaling, we introduce a normalized acceleration rate, $\mathcal{C}^{-1} \equiv d(\gamma\beta_{\rm sh}^{-2})/d(t\Omega_{\rm ci})$. We find a characteristic universal value of $\mathcal{C}^{-1} \simeq 0.1$, which varies by less than a factor of $2$ across our full survey (Section~\ref{subsec:proton_acceleration}). This rate exhibits almost no dependence on $M_s$ or $v_{\rm sh}$ and only a weak dependence on $T_i/T_e$. Modest reductions occur only at very low $M_A$ (weaker self-generated turbulence) or very high $M_A$ (disruption by intermittent magnetic structures).
    
    \item \textbf{Ion Energy Partition:} The downstream nonthermal ion energy fraction is maintained at $\sim 10\%$ of the total shock kinetic energy across essentially all explored parameters (Section~\ref{subsec:energy_partition}). Remarkably, efficient proton acceleration persists even at sonic Mach numbers as low as $M_s \simeq 2$, a regime previously considered unfavorable for strong nonthermal particle production.

    \item \textbf{Electron Acceleration Efficiency:} We find that electron acceleration is energetically subdominant in our survey. The nonthermal electron fraction is $\lesssim 1\%$, and is typically $\sim 10^{-3}$ for coronal-like $T_i/T_e$. This fraction decreases with increasing $T_i/T_e$. For several trans-relativistic runs with extreme magnetic field amplification, the electron cutoff energy saturates, suggesting that sustained electron DSA is inhibited when the amplified fields intermittently make the shock effectively superluminal for electrons (Section~\ref{subsection:disscus_electron}).

\end{enumerate}

These kinetic results provide strong physical support for the scenario where IceCube neutrinos originate from DSA-accelerated protons. Applying our derived normalized rate ($\mathcal{C}^{-1}\simeq 0.1$) to the specific conditions of NGC~1068, we demonstrate that protons can be accelerated to $\sim 100$~TeV on timescales shorter than the relevant cooling and dynamical limits (Section~\ref{subsection:NGC1068}). Furthermore, the nonthermal proton power required to explain the observed neutrino luminosity represents only a few percent of the estimated shock power---well within the $10\%$ efficiency limit found in our simulations. Thus, shocks in SMBH coronae can naturally satisfy both the spectral and energetic requirements for neutrino production, while accompanying gamma-rays are attenuated by the intense coronal radiation field.

\begin{acknowledgments}
This work was partly achieved through the use of SQUID at the D3 Center, The University of Osaka and HPE Cray XD2000 at the Center for Computational Astrophysics, National Astronomical Observatory of Japan. Numerical computations were also carried out on the ``Flow'' supercomputer at the Information Technology Center, Nagoya University through the HPCI System Research Project (Project ID: hp250093). This research was supported by JSPS KAKENHI grant Nos. JP24H00204, JP23H04864 and by JST SPRING, grants No. JPMJSP2138. T.S. acknowledges support from JSPS KAKENHI grants No. JP26K07043. Y.I. is supported by JSPS KAKENHI grant Nos. JP22K18277 and JP26H00604, and WPI, MEXT, Japan.
\end{acknowledgments}

\appendix

\section{Complete Ion and Electron Downstream Spectra} \label{sec:Appendix_A}
\begin{figure*}[h]
     \centering
     \includegraphics[width=0.99\textwidth]{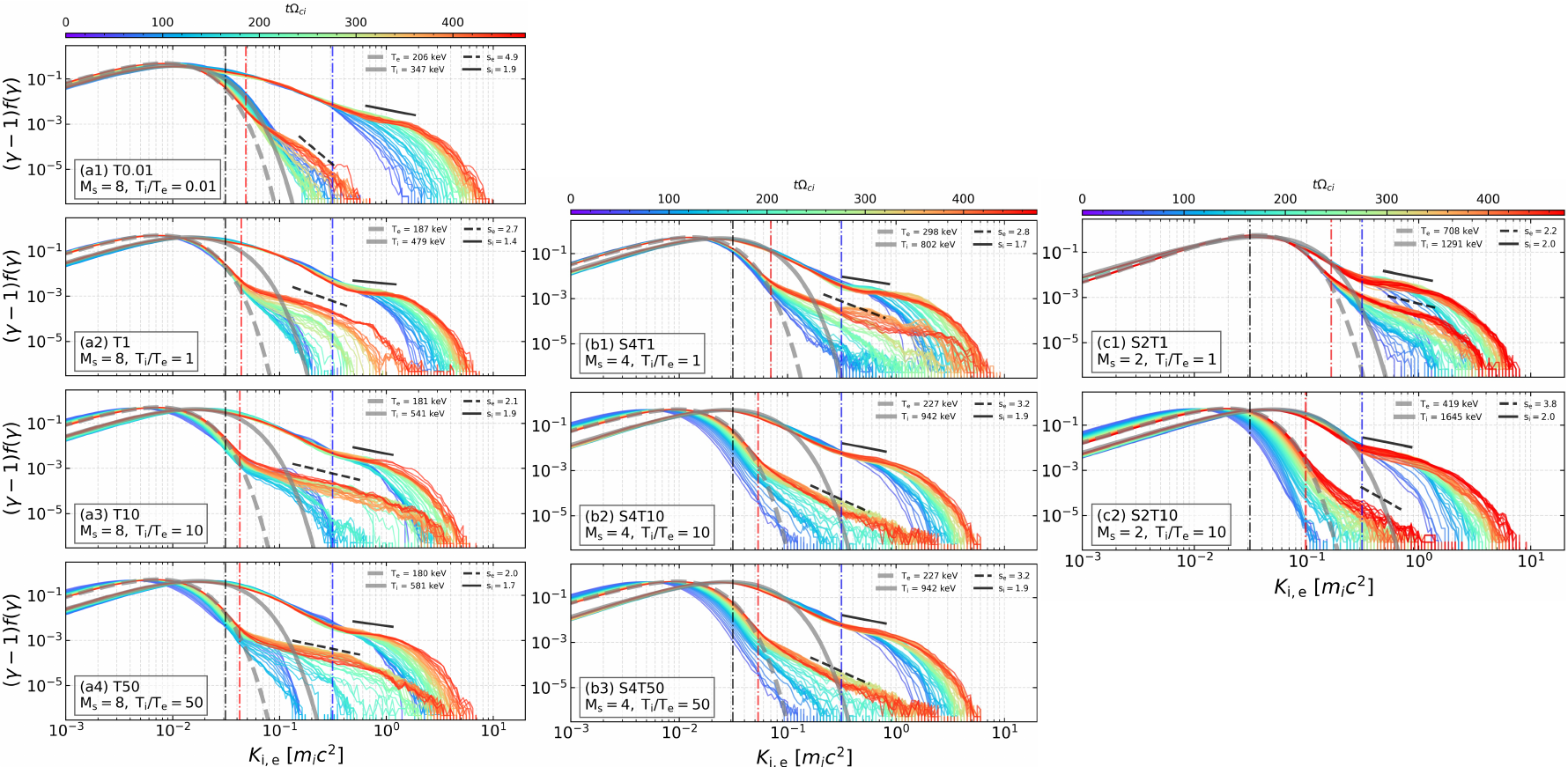}
\caption{\label{fig:spec_apdx1} Downstream spectra of ions and electrons for all simulations with fixed $M_A = 26$ and $v_{\rm pt}/c = 0.25$. Left panels (a1)--(a4) show cases with $M_s = 8$ while varying $T_i/T_e$ from 0.01 to 50 [runs T0.01, T1, T10 (reference run), and T50]. Middle panels (b1)--(b3) correspond to simulations with fixed $M_s = 4$ and $T_i/T_e = 1$--$50$ (runs S4T1, S4T10, and S4T50). Right panels (c1) and (c2) display simulations with $M_s = 2$ and $T_i/T_e = 1$ and $10$ (runs S2T1 and S2T10).}
\end{figure*}

In this appendix, we present the downstream ion and electron spectra for the full suite of simulations, following the format of the reference run shown in Figure~\ref{fig:spec_track}(a). Figure~\ref{fig:spec_apdx1} displays cases with fixed $M_A = 26$ and $v_{\rm pt}/c = 0.25$, but varying sonic Mach numbers and temperature ratios; here, we focus on the low sonic Mach number regime ($M_s \leq 8$).

\begin{figure*}[t]
     \centering
     \includegraphics[width=0.99\textwidth]{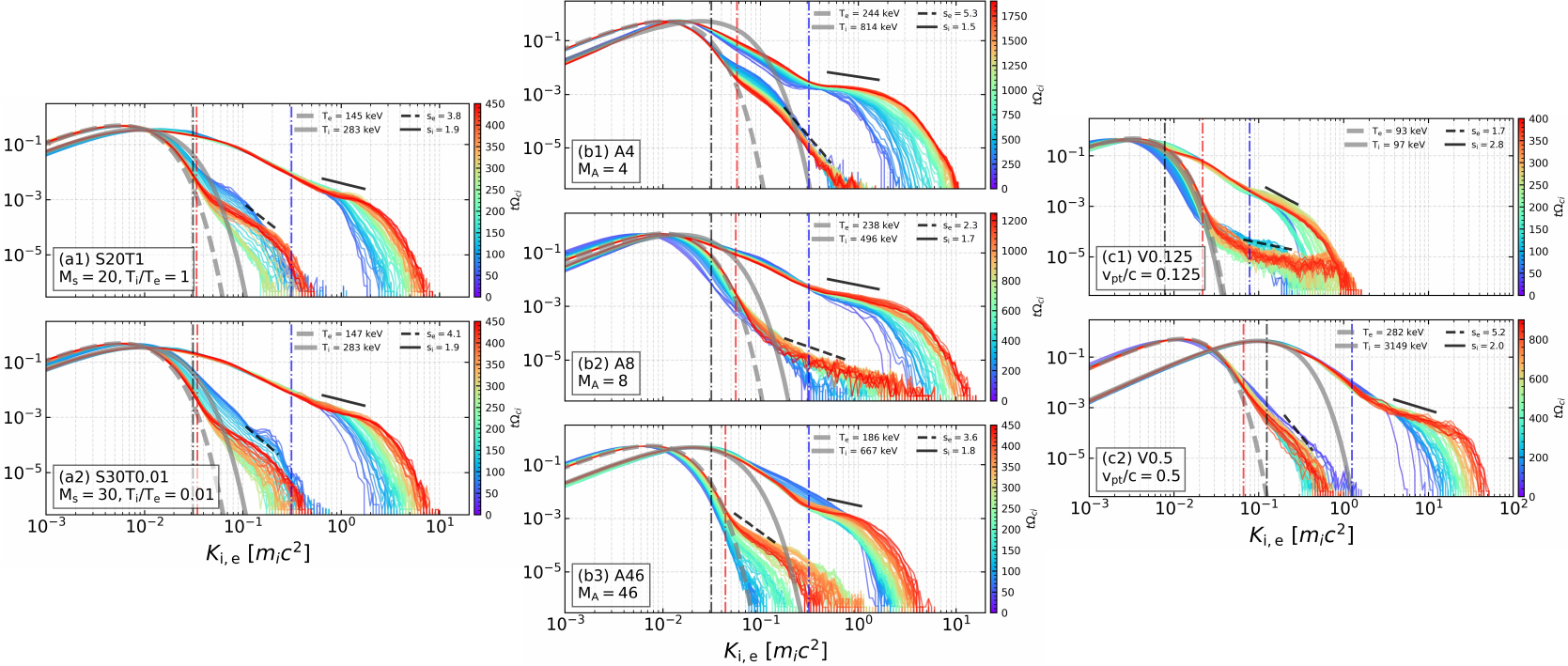}
\caption{\label{fig:spec_apdx_diffB} Downstream spectra of ions and electrons for simulations with high $M_s$, varying $M_A$, and varying $v_{\rm pt}$. Left panels, (a1) and (a2), show high sonic Mach number cases ($M_s = 20$ and $30$) with fixed $M_A = 26$ and $v_{\rm pt}/c = 0.25$, similar to Figure~\ref{fig:spec_apdx1}. The middle and right panels share fixed values of $M_s = 8$ and $T_i/T_e = 10$. Middle panels (b1)--(b3) correspond to simulations with fixed $v_{\rm pt}/c = 0.25$ and varying $M_A$ (runs A4, A8, and A46). Right panels (c1) and (c2) display varying $v_{\rm pt}/c$ ($0.125$ and $0.5$) with fixed $M_A$ (runs V0.125 and V0.5).}
\end{figure*}

High sonic Mach number cases (representing the cold inflow scenario described in Section~\ref{sec:condition_BHCS}) are presented in the left panels of Figure~\ref{fig:spec_apdx_diffB}. Additionally, the middle and right panels of Figure~\ref{fig:spec_apdx_diffB} display spectra for simulations with varying $M_A$ (at fixed $v_{\rm pt}/c = 0.25$) and varying $v_{\rm pt}/c$ (at fixed $M_A = 26$), respectively.

\bibliography{reference}{}

@article{kelner2006EnergySpectraGamma,
  title = {Energy spectra of gamma rays, electrons, and neutrinos produced at proton-proton interactions in the very high energy regime},
  volume = {74},
  url = {https://link.aps.org/doi/10.1103/PhysRevD.74.034018},
  doi = {10.1103/PhysRevD.74.034018},
  number = {3},
  journal = {Physical Review D},
  author = {Kelner, S. R. and Aharonian, F. A. and Bugayov, V. V.},
  year = {2006},
  pages = {034018}
}

@article{kelner2008EnergySpectraGamma,
  title = {Energy spectra of gamma rays, electrons, and neutrinos produced at interactions of relativistic protons with low energy radiation},
  volume = {78},
  url = {https://link.aps.org/doi/10.1103/PhysRevD.78.034013},
  doi = {10.1103/PhysRevD.78.034013},
  number = {3},
  journal = {Physical Review D},
  author = {Kelner, S. R. and Aharonian, F. A.},
  year = {2008},
  pages = {034013}
}

@article{icecubecollaboration*+2022EvidenceNeutrinoEmission,
  author = {Abbasi, R. and others},
  title = {Evidence for neutrino emission from the nearby active galaxy {NGC} 1068},
  journal = {Science},
  volume = {378},
  pages = {538--543},
  year = {2022},
  doi = {10.1126/science.abg3395}
}

@article{abbasi_evidence_2025,
  author = {Abbasi, R. and others},
  title = {Evidence for Neutrino Emission from X-ray Bright Active Galactic Nuclei with IceCube},
  journal = {Physical Review Letters},
  volume = {136},
  pages = {121002},
  year = {2025}
}

@article{yasuda2024NeutrinosGammaRays,
  title = {Neutrinos and {Gamma} {Rays} from {Beta} {Decays} in an {Active} {Galactic} {Nucleus} {NGC} 1068 {Jet}},
  volume = {134},
  url = {https://link.aps.org/doi/10.1103/PhysRevLett.134.151005},
  doi = {10.1103/PhysRevLett.134.151005},
  number = {15},
  journal = {Physical Review Letters},
  author = {Yasuda, Koichiro and Sakai, Nobuyuki and Inoue, Yoshiyuki and Kusenko, Alexander},
  year = {2025},
  pages = {151005}
}

@article{inoue2019HighenergyParticlesAccretion,
  title = {On {High}-energy {Particles} in {Accretion} {Disk} {Coronae} of {Supermassive} {Black} {Holes}: {Implications} for {MeV} {Gamma}-rays and {High}-energy {Neutrinos} from {AGN} {Cores}},
  volume = {880},
  url = {https://iopscience.iop.org/article/10.3847/1538-4357/ab2715},
  doi = {10.3847/1538-4357/ab2715},
  number = {1},
  journal = {The Astrophysical Journal},
  author = {Inoue, Yoshiyuki and Khangulyan, Dmitry and Inoue, Susumu and Doi, Akihiro},
  year = {2019},
  pages = {40}
}

@article{murase2020HiddenCoresActive,
  title = {Hidden {Cores} of {Active} {Galactic} {Nuclei} as the {Origin} of {Medium}-{Energy} {Neutrinos}: {Critical} {Tests} with the {MeV} {Gamma}-{Ray} {Connection}},
  volume = {125},
  url = {https://link.aps.org/doi/10.1103/PhysRevLett.125.011101},
  doi = {10.1103/PhysRevLett.125.011101},
  number = {1},
  journal = {Physical Review Letters},
  author = {Murase, Kohta and Kimura, Shigeo S. and Mészáros, Peter},
  year = {2020},
  pages = {011101}
}

@ARTICLE{kawabata2010PASJ...62..621K,
  author = {{Kawabata}, Ryoji and {Mineshige}, Shin},
  title = "{Radiative Spectra from Disk Corona and Inner Hot Flow in Black-Hole X-Ray Binaries}",
  journal = {\pasj},
  year = 2010,
  volume = {62},
  pages = {621},
  doi = {10.1093/pasj/62.3.621},
  archivePrefix = {arXiv},
  eprint = {1003.1430},
  primaryClass = {astro-ph.HE}
}

@article{yoshiyuki_inoue_origin_2020,
  author = {{Inoue}, Yoshiyuki and {Khangulyan}, Dmitry and {Doi}, Akihiro},
  title = "{On the Origin of High-energy Neutrinos from NGC 1068: The Role of Nonthermal Coronal Activity}",
  journal = {\apjl},
  year = 2020,
  volume = {891},
  number = {2},
  eid = {L33},
  pages = {L33},
  doi = {10.3847/2041-8213/ab7661},
  archivePrefix = {arXiv},
  eprint = {1909.02239},
  primaryClass = {astro-ph.HE}
}

@ARTICLE{Inoue2022arXiv220702097I,
  author = {{Inoue}, Susumu and {Cerruti}, Matteo and {Murase}, Kohta and {Liu}, Ruo-Yu},
  title = "{High-energy neutrinos and gamma rays from winds and tori in active galactic nuclei}",
  journal = {arXiv e-prints},
  year = 2022,
  eid = {arXiv:2207.02097},
  pages = {arXiv:2207.02097},
  doi = {10.48550/arXiv.2207.02097},
  archivePrefix = {arXiv},
  eprint = {2207.02097},
  primaryClass = {astro-ph.HE}
}

@article{kheirandish2021HighenergyNeutrinosMagnetized,
  title = {High-energy {Neutrinos} from {Magnetized} {Coronae} of {Active} {Galactic} {Nuclei} and {Prospects} for {Identification} of {Seyfert} {Galaxies} and {Quasars} in {Neutrino} {Telescopes}},
  volume = {922},
  url = {https://iopscience.iop.org/article/10.3847/1538-4357/ac1c77},
  doi = {10.3847/1538-4357/ac1c77},
  number = {1},
  journal = {The Astrophysical Journal},
  author = {Kheirandish, Ali and Murase, Kohta and Kimura, Shigeo S.},
  year = {2021},
  pages = {45}
}

@article{fiorillo2024MagnetizedStronglyTurbulenta,
  title = {A {Magnetized} {Strongly} {Turbulent} {Corona} as the {Source} of {Neutrinos} from {NGC} 1068},
  volume = {974},
  url = {https://iopscience.iop.org/article/10.3847/1538-4357/ad7021},
  doi = {10.3847/1538-4357/ad7021},
  number = {1},
  journal = {The Astrophysical Journal},
  author = {Fiorillo, Damiano F. G. and Comisso, Luca and Peretti, Enrico and Petropoulou, Maria and Sironi, Lorenzo},
  year = {2024},
  pages = {75}
}

@article{mbarek2024InterplayAcceleratedProtons,
  title = {Interplay between Accelerated Protons, x Rays and Neutrinos in the Corona of {{NGC}} 1068: {{Constraints}} from Kinetic Plasma Simulations},
  author = {Mbarek, Rostom and Philippov, Alexander and Chernoglazov, Alexander and Levinson, Amir and Mushotzky, Richard},
  year = 2024,
  journal = {Physical Review D},
  volume = {109},
  number = {10},
  pages = {L101306},
  publisher = {American Physical Society},
  doi = {10.1103/PhysRevD.109.L101306}
}

@article{Groselj2026arXiv260100518G,
  author = {Groselj, D. and Philippov, A. and Beloborodov, A. M. and Mushotzky, R.},
  title = {High-energy Emission from Turbulent Electron-ion Coronae of Accreting Black Holes},
  journal = {The Astrophysical Journal},
  volume = {1001},
  pages = {64},
  year = {2026}
}

@article{fiorillo2023TeVNeutrinosHard,
  title = {{TeV} {Neutrinos} and {Hard} {X}-{Rays} from {Relativistic} {Reconnection} in the {Corona} of {NGC} 1068},
  volume = {961},
  url = {https://iopscience.iop.org/article/10.3847/2041-8213/ad192b},
  doi = {10.3847/2041-8213/ad192b},
  number = {1},
  journal = {The Astrophysical Journal Letters},
  author = {Fiorillo, Damiano F. G. and Petropoulou, Maria and Comisso, Luca and Peretti, Enrico and Sironi, Lorenzo},
  year = {2024},
  note = {arXiv:2310.18254 [astro-ph]},
  pages = {L14}
}

@article{karavola2024NeutrinoPairCreation,
  author = {Karavola, D. and Petropoulou, M. and Fiorillo, D. F. G. and Comisso, L. and Sironi, L.},
  title = {Neutrino and pair creation in reconnection-powered coronae of accreting black holes},
  journal = {Journal of Cosmology and Astroparticle Physics},
  volume = {2025},
  pages = {075},
  year = {2025}
}

@article{lemoine_neutrinos_2025,
  title = {Neutrinos from stochastic acceleration in black hole environments},
  volume = {697},
  url = {https://www.aanda.org/10.1051/0004-6361/202453296},
  doi = {10.1051/0004-6361/202453296},
  journal = {Astronomy \& Astrophysics},
  author = {Lemoine, Martin and Rieger, Frank},
  year = {2025},
  note = {arXiv:2412.01457 [astro-ph]},
  pages = {A124}
}

@article{inoue2018DetectionCoronalMagnetica,
  title = {Detection of {Coronal} {Magnetic} {Activity} in nearby {Active} {Supermassive} {Black} {Holes}},
  volume = {869},
  url = {https://iopscience.iop.org/article/10.3847/1538-4357/aaeb95},
  doi = {10.3847/1538-4357/aaeb95},
  number = {2},
  journal = {The Astrophysical Journal},
  author = {Inoue, Yoshiyuki and Doi, Akihiro},
  year = {2018},
  pages = {114}
}

@ARTICLE{Michiyama2023PASJ...75..874M,
  author = {{Michiyama}, Tomonari and {Inoue}, Yoshiyuki and {Doi}, Akihiro},
  title = "{The centimeter-to-submillimeter broad-band radio spectrum of the central compact component in a nearby type-II Seyfert galaxy NGC 1068}",
  journal = {\pasj},
  year = 2023,
  volume = {75},
  number = {5},
  pages = {874-882},
  doi = {10.1093/pasj/psad044},
  archivePrefix = {arXiv},
  eprint = {2306.15950},
  primaryClass = {astro-ph.GA}
}

@ARTICLE{Michiyama2024ApJ...965...68M,
  author = {{Michiyama}, Tomonari and {Inoue}, Yoshiyuki and {Doi}, Akihiro and {Yamada}, Tomoya and {Fukazawa}, Yasushi and {Kubo}, Hidetoshi and {Barnier}, Samuel},
  title = "{ALMA Confirmation of Millimeter Time Variability in the Gamma-Ray Detected Seyfert Galaxy GRS 1734-292}",
  journal = {\apj},
  year = 2024,
  volume = {965},
  number = {1},
  eid = {68},
  pages = {68},
  doi = {10.3847/1538-4357/ad2fae},
  archivePrefix = {arXiv},
  eprint = {2404.00647},
  primaryClass = {astro-ph.GA}
}

@article{Shablovinskaya2024A&A...690A.232S,
  author = {Shablovinskaya, E. and Ricci, C. and Chang, C.-S. and others},
  title = {Joint ALMA/X-ray monitoring of the radio-quiet type 1 active galactic nucleus IC 4329A},
  journal = {Astronomy \& Astrophysics},
  volume = {690},
  pages = {A232},
  year = {2024},
  doi = {10.1051/0004-6361/202450133}
}

@article{palacio_millimeter_2025,
  author = {{del Palacio}, S. and Yang, C. and Aalto, S. and others},
  title = {Millimeter emission from supermassive black hole coronae},
  journal = {Astronomy \& Astrophysics},
  volume = {701},
  pages = {A41},
  year = {2025},
  doi = {10.1051/0004-6361/202554936}
}

@article{Jana2025A&A...699A..62J,
  author = {Jana, A. and Ricci, C. and Venselaar, S. M. and others},
  title = {ALMA observation of an evolving magnetized corona in the radio-quiet changing-state active galactic nucleus NGC 1566},
  journal = {Astronomy \& Astrophysics},
  volume = {699},
  pages = {A62},
  year = {2025},
  doi = {10.1051/0004-6361/202554491}
}

@article{mutie_consistent_2025,
  author = {Mutie, I. M. and {del Palacio}, S. and Beswick, R. J. and others},
  title = {A consistent radio to sub-mm pc-scale study of the nucleus of {NGC} 1068},
  journal = {Monthly Notices of the Royal Astronomical Society},
  volume = {539},
  pages = {808--819},
  year = {2025},
  doi = {10.1093/mnras/staf524}
}

@article{Hankla2025arXiv251201662H,
  author = {Hankla, A. and Philippov, A. and Mbarek, R. and others},
  title = {An outflow from the X-ray corona as the origin of millimeter emission from radio-quiet AGN},
  journal = {The Astrophysical Journal},
  volume = {997},
  pages = {224},
  year = {2025}
}

@article{ha_proton_2018,
  title = {Proton {Acceleration} in {Weak} {Quasi}-parallel {Intracluster} {Shocks}: {Injection} and {Early} {Acceleration}},
  volume = {864},
  url = {https://iopscience.iop.org/article/10.3847/1538-4357/aad634},
  doi = {10.3847/1538-4357/aad634},
  number = {2},
  journal = {The Astrophysical Journal},
  author = {Ha, Ji-Hoon and Ryu, Dongsu and Kang, Hyesung and Marle, Allard Jan Van},
  year = {2018},
  pages = {105}
}

@article{kang_electron_2019,
  title = {Electron {Preacceleration} in {Weak} {Quasi}-perpendicular {Shocks} in {High}-beta {Intracluster} {Medium}},
  volume = {876},
  url = {https://iopscience.iop.org/article/10.3847/1538-4357/ab16d1},
  doi = {10.3847/1538-4357/ab16d1},
  number = {1},
  journal = {The Astrophysical Journal},
  author = {Kang, Hyesung and Ryu, Dongsu and Ha, Ji-Hoon},
  year = {2019},
  pages = {79}
}

@article{vanmarle_influence_2020,
  author = {{van Marle}, A. J.},
  title = {On the influence of supra-thermal particle acceleration on the morphology of low-Mach, high-beta shocks},
  journal = {Monthly Notices of the Royal Astronomical Society},
  volume = {496},
  pages = {3198--3208},
  year = {2020},
  doi = {10.1093/mnras/staa1771}
}

@article{muller2020RadiationImpactBroadline,
  title = {Radiation from the impact of broad-line region clouds onto {AGN} accretion disks},
  volume = {636},
  url = {https://www.aanda.org/10.1051/0004-6361/202037639},
  doi = {10.1051/0004-6361/202037639},
  journal = {Astronomy \& Astrophysics},
  author = {Müller, A. L. and Romero, G. E.},
  year = {2020},
  pages = {A92}
}

@article{muller2022NonthermalEmissionFallback,
  title = {Nonthermal {Emission} from {Fall}-back {Clouds} in the {Broad}-line {Region} of {Active} {Galactic} {Nuclei}},
  volume = {931},
  url = {https://iopscience.iop.org/article/10.3847/1538-4357/ac660a},
  doi = {10.3847/1538-4357/ac660a},
  number = {1},
  journal = {The Astrophysical Journal},
  author = {Müller, Ana Laura and Naddaf, Mohammad-Hassan and Zajaček, Michal and Czerny, Bożena and Araudo, Anabella and Karas, Vladimír},
  year = {2022},
  pages = {39}
}

@article{sotomayor2022NonthermalRadiationCentral,
  title = {Nonthermal radiation from the central region of super-accreting active galactic nuclei},
  volume = {664},
  url = {https://www.aanda.org/10.1051/0004-6361/202243682},
  doi = {10.1051/0004-6361/202243682},
  journal = {Astronomy \& Astrophysics},
  author = {Sotomayor, Pablo and Romero, Gustavo E.},
  year = {2022},
  pages = {A178}
}

@article{fabian_properties_2015,
  title = {Properties of {AGN} coronae in the \textit{{NuSTAR}} era},
  volume = {451},
  url = {https://academic.oup.com/mnras/article-lookup/doi/10.1093/mnras/stv1218},
  doi = {10.1093/mnras/stv1218},
  number = {4},
  journal = {Monthly Notices of the Royal Astronomical Society},
  author = {Fabian, A. C. and Lohfink, A. and Kara, E. and Parker, M. L. and Vasudevan, R. and Reynolds, C. S.},
  year = {2015},
  pages = {4375--4383}
}

@article{laha_x-ray_2025,
  title = {X-ray properties of coronal emission in radio quiet active galactic nuclei},
  volume = {11},
  url = {https://www.frontiersin.org/articles/10.3389/fspas.2024.1530392/full},
  doi = {10.3389/fspas.2024.1530392},
  journal = {Frontiers in Astronomy and Space Sciences},
  author = {Laha, Sibasish and Ricci, Claudio and Mather, John C. and Behar, Ehud and Gallo, Luigi and Marin, Frederic and Mbarek, Rostom and Hankla, Amelia},
  year = {2025},
  pages = {1530392}
}

@article{yuan2014HotAccretionFlows,
  title = {Hot {Accretion} {Flows} {Around} {Black} {Holes}},
  volume = {52},
  url = {https://www.annualreviews.org/doi/10.1146/annurev-astro-082812-141003},
  doi = {10.1146/annurev-astro-082812-141003},
  number = {1},
  journal = {Annual Review of Astronomy and Astrophysics},
  author = {Yuan, Feng and Narayan, Ramesh},
  year = {2014},
  pages = {529--588}
}

@article{inoue2024UpperLimitCoronal,
  author = {Inoue, Y. and Takasao, S. and Khangulyan, D.},
  title = {Upper Limit on the Coronal Cosmic Ray Energy Budget in Seyfert Galaxies},
  journal = {Publications of the Astronomical Society of Japan},
  volume = {76},
  pages = {996},
  year = {2024}
}

@article{padovani2024HighenergyNeutrinosVicinity,
  author = {Padovani, P. and Resconi, E. and Ajello, M. and others},
  title = {High-energy neutrinos from the vicinity of the supermassive black hole in {NGC} 1068},
  journal = {Nature Astronomy},
  volume = {8},
  pages = {1077},
  year = {2024}
}

@article{murase2022HiddenHeartsNeutrino,
  title = {Hidden {Hearts} of {Neutrino} {Active} {Galaxies}},
  volume = {941},
  url = {https://iopscience.iop.org/article/10.3847/2041-8213/aca53c},
  doi = {10.3847/2041-8213/aca53c},
  number = {1},
  journal = {The Astrophysical Journal Letters},
  author = {Murase, Kohta},
  year = {2022},
  pages = {L17}
}

@article{das2024RevealingProductionMechanism,
  title = {Revealing the {Production} {Mechanism} of {High}-energy {Neutrinos} from {NGC} 1068},
  volume = {972},
  url = {https://iopscience.iop.org/article/10.3847/1538-4357/ad5a04},
  doi = {10.3847/1538-4357/ad5a04},
  number = {1},
  journal = {The Astrophysical Journal},
  author = {Das, Abhishek and Zhang, B. Theodore and Murase, Kohta},
  year = {2024},
  pages = {44}
}

@article{derouillat_smilei_2018,
  title = {Smilei : {A} collaborative, open-source, multi-purpose particle-in-cell code for plasma simulation},
  volume = {222},
  url = {https://linkinghub.elsevier.com/retrieve/pii/S0010465517303314},
  doi = {10.1016/j.cpc.2017.09.024},
  journal = {Computer Physics Communications},
  author = {Derouillat, J. and Beck, A. and Pérez, F. and Vinci, T. and Chiaramello, M. and Grassi, A. and Flé, M. and Bouchard, G. and Plotnikov, I. and Aunai, N. and Dargent, J. and Riconda, C. and Grech, M.},
  year = {2018},
  pages = {351--373}
}

@article{park2015SimultaneousAccelerationProtons,
  title = {Simultaneous {Acceleration} of {Protons} and {Electrons} at {Nonrelativistic} {Quasiparallel} {Collisionless} {Shocks}},
  volume = {114},
  url = {https://link.aps.org/doi/10.1103/PhysRevLett.114.085003},
  doi = {10.1103/PhysRevLett.114.085003},
  number = {8},
  journal = {Physical Review Letters},
  author = {Park, Jaehong and Caprioli, Damiano and Spitkovsky, Anatoly},
  year = {2015},
  pages = {085003}
}

@article{shalaby2024EnergyDissipationStrong,
  title = {Energy {Dissipation} in {Strong} {Collisionless} {Shocks}: {The} {Crucial} {Role} of {Ion}-to-electron {Scale} {Separation} in {Particle}-in-cell {Simulations}},
  volume = {977},
  url = {https://iopscience.iop.org/article/10.3847/2041-8213/ad99d8},
  doi = {10.3847/2041-8213/ad99d8},
  number = {2},
  journal = {The Astrophysical Journal Letters},
  author = {Shalaby, Mohamad},
  year = {2024},
  pages = {L43}
}

@article{gupta2024ReturnCurrentsCollisionless,
  title = {Return {Currents} in {Collisionless} {Shocks}},
  volume = {968},
  url = {https://iopscience.iop.org/article/10.3847/1538-4357/ad3e75},
  doi = {10.3847/1538-4357/ad3e75},
  number = {1},
  journal = {The Astrophysical Journal},
  author = {Gupta, Siddhartha and Caprioli, Damiano and Spitkovsky, Anatoly},
  year = {2024},
  pages = {17}
}

@article{gupta2025SpeeddependentThresholdElectron,
  author = {Gupta, S. and Caprioli, D. and Spitkovsky, A.},
  title = {Speed-dependent Threshold for Electron Injection into Diffusive Shock Acceleration},
  journal = {The Astrophysical Journal Letters},
  volume = {994},
  pages = {L34},
  year = {2025}
}

@article{park_particle--cell_2012,
  title = {Particle-in-cell simulations of particle energization from low {Mach} number fast mode shocks},
  volume = {19},
  url = {https://pubs.aip.org/pop/article/19/6/062904/381150/Particle-in-cell-simulations-of-particle},
  doi = {10.1063/1.4729913},
  number = {6},
  journal = {Physics of Plasmas},
  author = {Park, Jaehong and Workman, Jared C. and Blackman, Eric G. and Ren, Chuang and Siller, Robert},
  year = {2012},
  pages = {062904}
}

@article{caprioli2014SIMULATIONSIONACCELERATION,
  title = {{SIMULATIONS} {OF} {ION} {ACCELERATION} {AT} {NON}-{RELATIVISTIC} {SHOCKS}. {I}. {ACCELERATION} {EFFICIENCY}},
  volume = {783},
  url = {https://iopscience.iop.org/article/10.1088/0004-637X/783/2/91},
  doi = {10.1088/0004-637X/783/2/91},
  number = {2},
  journal = {The Astrophysical Journal},
  author = {Caprioli, D. and Spitkovsky, A.},
  year = {2014},
  pages = {91}
}

@article{haggerty2019DHybridRHybridParticleinCellCode,
  title = {{dHybridR}: a {Hybrid}--{Particle}-in-{Cell} {Code} {Including} {Relativistic} {Ion} {Dynamics}},
  volume = {887},
  url = {http://arxiv.org/abs/1909.05255},
  doi = {10.3847/1538-4357/ab58c8},
  number = {2},
  journal = {The Astrophysical Journal},
  author = {Haggerty, Colby C. and Caprioli, Damiano},
  year = {2019},
  note = {arXiv:1909.05255 [astro-ph]},
  pages = {165}
}

@article{kato2015PARTICLEACCELERATIONWAVE,
  title = {{PARTICLE} {ACCELERATION} {AND} {WAVE} {EXCITATION} {IN} {QUASI}-{PARALLEL} {HIGH}-{MACH}-{NUMBER} {COLLISIONLESS} {SHOCKS}: {PARTICLE}-{IN}-{CELL} {SIMULATION}},
  volume = {802},
  url = {https://iopscience.iop.org/article/10.1088/0004-637X/802/2/115},
  doi = {10.1088/0004-637X/802/2/115},
  number = {2},
  journal = {The Astrophysical Journal},
  author = {Kato, Tsunehiko N.},
  year = {2015},
  pages = {115}
}

@article{marcowith2021CosmicRaydrivenStreaming,
  title = {The cosmic ray-driven streaming instability in astrophysical and space plasmas},
  volume = {28},
  url = {https://pubs.aip.org/pop/article/28/8/080601/106688/The-cosmic-ray-driven-streaming-instability-in},
  doi = {10.1063/5.0013662},
  number = {8},
  journal = {Physics of Plasmas},
  author = {Marcowith, A. and Van Marle, A. J. and Plotnikov, I.},
  year = {2021},
  pages = {080601}
}

@article{skilling_cosmic_1971,
  title = {Cosmic {Rays} in the {Galaxy}: {Convection} or {Diffusion}?},
  volume = {170},
  url = {http://adsabs.harvard.edu/doi/10.1086/151210},
  doi = {10.1086/151210},
  journal = {The Astrophysical Journal},
  author = {Skilling, John},
  year = {1971},
  pages = {265}
}

@article{bell2004TurbulentAmplificationMagnetic,
  title = {Turbulent amplification of magnetic field and diffusive shock acceleration of cosmic rays},
  volume = {353},
  url = {https://academic.oup.com/mnras/article-lookup/doi/10.1111/j.1365-2966.2004.08097.x},
  doi = {10.1111/j.1365-2966.2004.08097.x},
  number = {2},
  journal = {Monthly Notices of the Royal Astronomical Society},
  author = {Bell, A. R.},
  year = {2004},
  pages = {550--558}
}

@article{gupta2024ElectronAccelerationQuasiparallel,
  title = {Electron {Acceleration} at {Quasi}-parallel {Nonrelativistic} {Shocks}: {A} {1D} {Kinetic} {Survey}},
  volume = {976},
  url = {https://iopscience.iop.org/article/10.3847/1538-4357/ad7c4c},
  doi = {10.3847/1538-4357/ad7c4c},
  number = {1},
  journal = {The Astrophysical Journal},
  author = {Gupta, Siddhartha and Caprioli, Damiano and Spitkovsky, Anatoly},
  year = {2024},
  pages = {10}
}

@article{caprioli2014SIMULATIONSIONACCELERATIONa,
  title = {{SIMULATIONS} {OF} {ION} {ACCELERATION} {AT} {NON}-{RELATIVISTIC} {SHOCKS}. {II}. {MAGNETIC} {FIELD} {AMPLIFICATION}},
  volume = {794},
  url = {https://iopscience.iop.org/article/10.1088/0004-637X/794/1/46},
  doi = {10.1088/0004-637X/794/1/46},
  number = {1},
  journal = {The Astrophysical Journal},
  author = {Caprioli, D. and Spitkovsky, A.},
  year = {2014},
  pages = {46}
}

@article{drury_introduction_1983,
  title = {An introduction to the theory of diffusive shock acceleration of energetic particles in tenuous plasmas},
  volume = {46},
  url = {https://iopscience.iop.org/article/10.1088/0034-4885/46/8/002},
  doi = {10.1088/0034-4885/46/8/002},
  number = {8},
  journal = {Reports on Progress in Physics},
  author = {Drury, L O'C},
  year = {1983},
  pages = {973--1027}
}

@article{blandford_particle_1987,
  title = {Particle acceleration at astrophysical shocks: {A} theory of cosmic ray origin},
  volume = {154},
  url = {https://linkinghub.elsevier.com/retrieve/pii/0370157387901347},
  doi = {10.1016/0370-1573(87)90134-7},
  number = {1},
  journal = {Physics Reports},
  author = {Blandford, Roger and Eichler, David},
  year = {1987},
  pages = {1--75}
}

@ARTICLE{vink_critical_2013,
  author = {{Vink}, Jacco and {Yamazaki}, Ryo},
  title = "{A Critical Shock Mach Number for Particle Acceleration in the Absence of Pre-existing Cosmic Rays: M=\textbackslashsqrt\{5\}}",
  journal = {\apj},
  year = 2014,
  volume = {780},
  number = {2},
  eid = {125},
  pages = {125},
  doi = {10.1088/0004-637X/780/2/125},
  archivePrefix = {arXiv},
  eprint = {1307.4754},
  primaryClass = {astro-ph.HE}
}

@article{zekovic_slams-propelled_2025,
  title = {{SLAMS}-propelled {Electron} {Acceleration} at {High}-{Mach}-number {Astrophysical} {Shocks}},
  volume = {988},
  url = {https://iopscience.iop.org/article/10.3847/1538-4357/ade05f},
  doi = {10.3847/1538-4357/ade05f},
  number = {1},
  journal = {The Astrophysical Journal},
  author = {Zeković, Vladimir and Spitkovsky, Anatoly and Hemler, Zachary},
  year = {2025},
  note = {arXiv:2408.02084 [astro-ph]},
  pages = {40}
}

@article{shakura1973BlackHolesBinary,
  title = {Black holes in binary systems. {Observational} appearance.},
  volume = {24},
  url = {https://ui.adsabs.harvard.edu/abs/1973A&A....24..337S},
  journal = {Astronomy and Astrophysics},
  author = {Shakura, N. I. and Sunyaev, R. A.},
  year = {1973},
  note = {ADS Bibcode: 1973A\&A....24..337S},
  pages = {337--355}
}

@article{hillas_origin_1984,
  title = {The {Origin} of {Ultra}-{High}-{Energy} {Cosmic} {Rays}},
  volume = {22},
  url = {https://www.annualreviews.org/doi/10.1146/annurev.aa.22.090184.002233},
  doi = {10.1146/annurev.aa.22.090184.002233},
  number = {1},
  journal = {Annual Review of Astronomy and Astrophysics},
  author = {Hillas, A. M.},
  year = {1984},
  pages = {425--444}
}

@article{drury2012OriginCosmicRays,
  title = {Origin of cosmic rays},
  volume = {39-40},
  url = {https://linkinghub.elsevier.com/retrieve/pii/S092765051200045X},
  doi = {10.1016/j.astropartphys.2012.02.006},
  journal = {Astroparticle Physics},
  author = {Drury, Luke O’C.},
  year = {2012},
  pages = {52--60}
}

@article{oka2025MaximumEnergyParticles,
  title = {Maximum {Energy} of {Particles} in {Plasmas}},
  volume = {979},
  url = {https://iopscience.iop.org/article/10.3847/1538-4357/ad9916},
  doi = {10.3847/1538-4357/ad9916},
  number = {2},
  journal = {The Astrophysical Journal},
  author = {Oka, Mitsuo and Makishima, Kazuo and Terasawa, Toshio},
  year = {2025},
  pages = {161}
}

@article{jikei_magnetic_2025,
  author = {Jikei, T. and Groselj, D. and Sironi, L.},
  title = {Magnetic Field Amplification and Particle Acceleration in Weakly Magnetized Trans-relativistic Electron-ion Shocks},
  journal = {The Astrophysical Journal},
  volume = {998},
  pages = {149},
  year = {2025}
}

@article{pcrumley2019KineticSimulationsMildly,
  author = {Crumley, P. and Caprioli, D. and Markoff, S. and Spitkovsky, A.},
  title = {Kinetic simulations of mildly relativistic shocks -- I. Particle acceleration in high Mach number shocks},
  journal = {Monthly Notices of the Royal Astronomical Society},
  volume = {485},
  pages = {5105--5119},
  year = {2019},
  doi = {10.1093/mnras/stz232}
}

@ARTICLE{Kawazura2019,
  author = {{Kawazura}, Yohei and {Barnes}, Michael and {Schekochihin}, Alexander A.},
  title = "{Thermal disequilibration of ions and electrons by collisionless plasma turbulence}",
  journal = {Proceedings of the National Academy of Science},
  year = 2019,
  volume = {116},
  number = {3},
  pages = {771-776},
  doi = {10.1073/pnas.1812491116},
  archivePrefix = {arXiv},
  eprint = {1807.07702},
  primaryClass = {physics.plasm-ph}
}

@article{gorbunov2024FirstprinciplesMeasurementIon,
  title = {First-principles {Measurement} of {Ion} and {Electron} {Energization} in {Collisionless} {Accretion} {Flows}},
  volume = {982},
  url = {https://iopscience.iop.org/article/10.3847/2041-8213/adbca4},
  doi = {10.3847/2041-8213/adbca4},
  number = {1},
  journal = {The Astrophysical Journal Letters},
  author = {Gorbunov, Evgeny A. and Bacchini, Fabio and Zhdankin, Vladimir and Werner, Gregory R. and Begelman, Mitchell C. and Uzdensky, Dmitri A.},
  year = {2025},
  pages = {L28}
}
\bibliographystyle{aasjournalv7}

\end{document}